\documentclass[a4paper,UKenglish]{new_tlp}

\pdfoutput=1

\usepackage{xcolor}
\usepackage{float}
\usepackage{microtype}
\usepackage{xspace}
\usepackage{url}
\usepackage{graphicx}
\usepackage[vlined,linesnumbered]{algorithm2e}
\usepackage{amssymb}
\usepackage{amsmath} 
\usepackage{tikz}
\usetikzlibrary{snakes,arrows,shapes}

\usepackage{listings}
\usepackage{dsfont}
\usepackage[draft]{fixme}
\usepackage{multicol}
\usepackage{paralist}
\usepackage{multirow}
\usepackage{lscape}

\def\llvm2kittle{\texttt{llvm2KITTeL}\xspace}
\def\koat{\texttt{KoAT}\xspace}
\def\cofloco{\texttt{CoFloCo}\xspace}
\def\pubs{\texttt{PUBs}\xspace}

\def\rankfinder{\texttt{iRankFinder}\xspace}
\def\irankfinder{\rankfinder}
\def\pe{\texttt{PE}\xspace}

\newcommand{\lst}[1]{\mbox{\lstinline{#1}}}

\usepackage{listings}
\newcommand{\tuple}[1]{\langle{#1}\rangle}
\newcommand{\node}[1]{\ensuremath{\mathtt{#1}}}

\newcommand{\edge}[3]{\ensuremath{{#1}\stackrel{#2}{\rightarrow}{#3}}\xspace}
\newcommand{\nodes}[1]{\ensuremath{\mathtt{nodes}(#1)}}

\newcommand{\trsys}[0]{\ensuremath{{\cal T}}\xspace}
\def\ll{[\![}
\def\rr{]\!]}
\newcommand{\sols}[1]{\ensuremath{\ll{#1}\rr}\xspace}

\newcommand{\Z}{\mathds{Z}}
\newcommand{\R}{\mathds{R}}

\newcommand{\its}[0]{ITS\xspace}
\newcommand{\itss}[0]{ITSs\xspace}
\newcommand{\scc}[0]{SCC\xspace}
\newcommand{\sccs}[0]{SCCs\xspace}

\newcommand{\cfr}[0]{CFR\xspace}
\newcommand{\lrf}[0]{LRF\xspace}
\newcommand{\lrfs}[0]{LRFs\xspace}
\newcommand{\llrf}[0]{LLRF\xspace}
\newcommand{\llrfs}[0]{LLRFs\xspace}
\newcommand{\mlrf}[0]{M{\ensuremath{\Phi}}RF\xspace}
\newcommand{\mlrfs}[0]{M{\ensuremath{\Phi}}RFs\xspace}
\newcommand{\chc}[0]{CHC\xspace}
\newcommand{\chcs}[0]{CHCs\xspace}

\newcommand{\crss}[0]{CRSs\xspace}

\newcommand{\cfra}[0]{\ensuremath{\mathtt{CFR_A}}\xspace}
\newcommand{\cfrb}[0]{\ensuremath{\mathtt{CFR_B}}\xspace}
\newcommand{\cfrs}[0]{\ensuremath{\mathtt{CFR_S}}\xspace}

\newcommand{\chct}[1]{\ensuremath{\mathtt{HC}({#1})}\xspace}

\newcommand{\props}[2]{\ensuremath{\mathtt{PR_{#1}^{#2}}}}
\newcommand{\propsh}[0]{\props{h}{}\xspace}
\newcommand{\propsc}[0]{\props{c}{}\xspace}
\newcommand{\propshv}[0]{\props{hv}{}\xspace}
\newcommand{\propscv}[0]{\props{cv}{}\xspace}
\newcommand{\propsdh}[0]{\props{h}{d}\xspace}

\newcommand{\proj}[2]{\ensuremath{\mathtt{proj}_{#1}(#2)}\xspace}
\newcommand{\nc}[1]{\ensuremath{\mathtt{ID}({#1})}\xspace}

\newcommand{\ebox}[0]{\hfill\ensuremath{\Box}}

\newcommand{\bsetA}[0]{\texttt{SET-1}\xspace}
\newcommand{\bsetB}[0]{\texttt{SET-2}\xspace}
\newcommand{\bsetC}[0]{\texttt{SET-3}\xspace}

\newcommand{\cbox}[1]{\colorbox{yellow}{\textcolor{blue}{#1}}}

\SetVlineSkip{0pt}      
\SetKwProg{Fn}{}{}{}
\SetKwFunction{proccfr}{{\sc PE}}
\SetKwFunction{proctermin}{{\sc Termin}}
\SetKwFunction{proctermincfg}{{\sc TerminITS}}
\SetKwFunction{procterminscc}{{\sc TerminSCC}}
\SetKwFunction{procremovenrn}{{\sc DelNonReaching}}
\SetKwFunction{procremovetn}{{\sc DelTerminating}}
\SetKwFunction{procbuildits}{{\sc ConITS}}

\newtheorem{example}{Example}[section]

\newsavebox{\fmbox}

\title[Control-Flow Refinement by Partial Evaluation]{Control-Flow Refinement by Partial Evaluation, and its Application to Termination and Cost Analysis%
\thanks{This work was funded partially by the Spanish MICINN/FEDER, UE project RTI2018-094403-B-C31, the MINECO project TIN2015-69175-C4-2-R, the CM project S2018/TCS-4314 and by the pre-doctoral UCM grant CT27/16-CT28/16.}
}

\author[Jes\'us J. Dom\'enech et. al.]{
Jes\'us J. Dom\'enech\\
Universidad Complutense de Madrid, Spain\\
\texttt{jdomenech@ucm.es}
\and
John P. Gallagher\\
Roskilde University, Denmark and IMDEA Software Institute, Spain\\
\texttt{jpg@ruc.dk}
\and
Samir Genaim\\
Universidad Complutense de Madrid, Spain\\
\texttt{sgenaim@ucm.es}
}

\begin{document}

\maketitle

\begin{abstract}
Control-flow refinement refers to program transformations whose purpose is to make implicit control-flow explicit, and is used in the context of program analysis to increase precision. Several techniques have been suggested for different programming models, typically tailored to improving precision for a particular analysis.
In this paper we explore the use of partial evaluation of Horn clauses as a general-purpose technique for control-flow refinement for integer transitions systems. These are control-flow graphs where edges are annotated with linear constraints describing transitions between corresponding nodes, and they are used in many program analysis tools. Using partial evaluation for control-flow refinement has the clear advantage over other approaches in that soundness follows from the general properties of partial evaluation; in particular,  properties such as termination and complexity are preserved. 
We use a partial evaluation algorithm incorporating property-based abstraction, and show how the right choice of properties allows us to prove termination and to infer complexity of challenging programs that cannot be handled by state-of-the-art tools.
We report on the integration of the technique in a termination analyzer, and its use as a preprocessing step for several cost analyzers.
\end{abstract}

\begin{keywords}
  Control-flow refinement, partial evaluation, termination analysis, cost analysis
\end{keywords}

\section{Introduction}
\label{sec:intro}

Control-flow refinement (\cfr) is used in
program analysis to make the implicit control-flow of a given program
explicit.
Consider for example the program on the left:

\begin{minipage}{5.3cm}
\begin{lstlisting}[numbers=none]
while ( x > 0 )
  if ( y < z ) y++; else x--;
\end{lstlisting}
\end{minipage}
~~~~
\begin{minipage}{5.3cm}
\begin{lstlisting}[numbers=none]
while (x > 0 && y < z) y++;
while (x > 0 && y >= z) x--;
\end{lstlisting}
\end{minipage}

\noindent
Its execution has two implicit phases: in the first one,  $\mathtt{y}$
is incremented until it reaches the value of $\mathtt{z}$, and in the second
phase $\mathtt{x}$ is decremented until it reaches $\mathtt{0}$.
\cfr techniques transform this program into the one on the right, in
which the two phases are explicit.
For termination analysis, the transformation
simplifies the termination proof; while the original program requires
a lexicographic termination argument, the transformed one requires
only linear ones. There are also cases where it is not possible to
prove termination without such transformations.
For cost analysis, tools based on bounding loop
iterations using linear ranking functions fail to infer the cost of
the first program, while a linear
upper-bound can be inferred for the second.
\cfr also helps in inferring more precise invariants, without the need
for 
disjunctive abstract domains.

\citeN{GulwaniJK09} and \citeN{Flores-MontoyaH14} used \cfr to improve
the precision of cost analysis. Roughly, they explore different
combinations of the paths of a given loop in order to discover
execution patterns, and then transform the loop to follow these
patterns.
\citeN{SharmaDDA11} used \cfr to improve invariants in order to prove
assertions. Their technique is based on finding a \emph{splitter
  predicate} such that when it holds one part of the loop is executed
and when it does not hold another part is executed. The loop is then
rewritten as two consecutive loops, where the predicate is required to
hold in one and not to hold in the other. For the loop above, $y<z$ is
a splitter predicate.

Since \cfr is, in principle, a program transformation that specializes
a programs to distinguish different execution scenarios, a natural
question to ask is whether such a specialization can be achieved by
partial evaluation, which is a general-purpose specialization
technique. Using partial evaluation for \cfr has the advantage that
soundness comes for free, and that it is not tailored for a particular
purpose but rather can be tuned depending on the application domain.
In this paper we apply partial evaluation for \cfr of integer
transition systems, which are control-flow graphs where edges are
annotated with formulas describing transitions, and can be represented
as Horn clause programs. This is a very popular model that is often
used in static analysis, and thus many tools would benefit from a \cfr
procedure for this setting.
We use the partial evaluation technique
of~\citeN{Gallagher19} that specializes Horn clause programs with respect to a
set of predefined properties.
For example, partial evaluation of the loop discussed above using the
properties $x>0$ and $y<z$ automatically discovers the two
phases. Note that the right choice of properties is crucial for
achieving the desired refinements.
The contributions are:
(1) we suggest heuristics for inferring properties for partial
evaluation automatically;
(2) we discuss the use of \cfr for termination analysis, using an
algorithm that applies \cfr only when needed, and for cost analysis as
a preprocessing step;
(3) we suggest the use of termination witnesses (ranking functions) as
properties for \cfr in order to improve precision of cost analysis;
and
(4) we provide a publicly available implementation and corresponding
experimental evaluation that demonstrates the usefulness of our \cfr
procedure.

Sect.~\ref{sec:prelim} gives some necessary background;
Sect.~\ref{sec:pe} describes a \cfr procedure that is based on using
partial evaluation; Sect.~\ref{sec:termin} discusses the use of our
\cfr procedure in the context of termination and cost analyses;
Sect.~\ref{sec:imp} discusses an implementation and experimental
evaluation, and Sect.~\ref{sec:conc} concludes and discusses related
and future work.

\begin{figure}[H]

\hrule
\smallskip

\vspace{-0.1cm}
\begin{minipage}{0.46\textwidth}
  \begin{lstlisting}
void phases1(int x, int y, int z) {
  while( x > 0 ) {
    if ( y < z ) y = y + 1;
    else x = x - 1;
  }
}
\end{lstlisting}
\vspace*{-0.45cm}
\[
\setlength\abovedisplayskip{2pt}%
\setlength\belowdisplayskip{2pt}%
\begin{array}{r@{}l}
\varphi_0\equiv&  \nc{x,y,z} \\
\varphi_1\equiv&  x>0 {\wedge} \nc{x,y,z} \\
\varphi_2\equiv&  x\leq 0 {\wedge} \nc{x,y,z} \\
\varphi_3\equiv&  y < z {\wedge}  y'=y+1 {\wedge} \nc{x,z} \\
\varphi_4\equiv&  y \geq z {\wedge} x'=x-1 {\wedge} \nc{y,z} \\
\end{array}
\]
\end{minipage}
~
\begin{minipage}{0.5\textwidth}
\begin{minipage}{0.3\textwidth}
  \fbox{
    \scalebox{.477}{%

\begin{tikzpicture}[>=latex,line join=bevel,]
"\Large"%
  \node (n0) at (12.5bp,131.0bp) [draw,fill=green,circle] {\scalebox{1.2}{$\node{n_0}$}};
  \node (n1) at (12.5bp,70.0bp) [draw,circle] {\scalebox{1.2}{$\node{n_1}$}};
  \node (n2) at (78.5bp,70.0bp) [draw,circle] {\scalebox{1.2}{$\node{n_2}$}};
  \node (n3) at (12.5bp,9.0bp) [draw,circle] {\scalebox{1.2}{$\node{n_3}$}};
  \node (trs) at (78.5bp,9.0bp) [] {\scalebox{1.2}{\cbox{$\trsys$}}};
  \path (n0) edge [->] node[left] {\scalebox{1.2}{$\varphi_0$}} (n1);
  \path (n1) edge [->] node[above] {\scalebox{1.2}{$\varphi_1$}} (n2);
  \path (n1) edge [->] node[left] {\scalebox{1.2}{$\varphi_2$}} (n3);
  \draw [->] (n2) ..controls (68.895bp,57.1bp) and (63.379bp,51.598bp)  .. (57.0bp,49.0bp) .. controls (47.533bp,45.144bp) and (43.467bp,45.144bp)  .. (34.0bp,49.0bp) .. controls (30.91bp,50.259bp) and (28.023bp,51.198bp)  .. (n1);
  \draw (45.5bp,58.0bp) node {\scalebox{1.2}{$\varphi_3$}};
  \draw [->] (n2) ..controls (68.895bp,82.9bp) and (63.379bp,88.402bp)  .. (57.0bp,91.0bp) .. controls (47.533bp,94.86bp) and (43.467bp,94.86bp)  .. (34.0bp,91.0bp) .. controls (30.91bp,89.74bp) and (28.023bp,87.802bp)  .. (n1);
  \draw (45.5bp,106.0bp) node {\scalebox{1.2}{$\varphi_4$}};
\end{tikzpicture}
}
  }
\end{minipage}
~~
\begin{minipage}{0.64\textwidth}
  \fbox{
    \scalebox{.4}{%

\begin{tikzpicture}[>=latex,line join=bevel,]
\Large%
  \node (n0) at (14.5bp,159.89bp) [draw,fill=green,circle] {\scalebox{1.3}{$\node{n_0}$}};
  \node (n1_3) at (78.5bp,159.89bp) [draw,circle] {\scalebox{1.3}{$\node{n_1^3}$}};
  \node (n3_2) at (144.5bp,159.89bp) [draw,circle] {\scalebox{1.3}{$\node{n_3^2}$}};

  \node (n2_2) at (78.5bp,91.89bp) [draw,circle] {\scalebox{1.3}{$\node{n_2^2}$}};
  \node (n1_2) at (12.5bp,91.89bp) [draw,circle] {\scalebox{1.3}{$\node{n_1^2}$}};
  \node (trs) at (144.5bp,91.89bp) [] {\scalebox{1.3}{\cbox{$\trsys_{pe}$}}};

  \node (n2_1) at (12.5bp,23.892bp) [draw,circle] {\scalebox{1.3}{$\node{n_2^1}$}};
  \node (n3_1) at (144.5bp,23.892bp) [draw,circle] {\scalebox{1.3}{$\node{n_3^1}$}};
  \node (n1_1) at (78.5bp,23.892bp) [draw,circle] {\scalebox{1.3}{$\node{n_1^1}$}};

  \path (n0) edge [->] node[above] {\scalebox{1.3}{$\varphi_0$}} (n1_3);
  \path (n1_3) edge [->] node[right] {\scalebox{1.3}{$\varphi_1$}} (n2_2);
  \path (n1_3) edge [->] node[above] {\scalebox{1.3}{$\varphi_2$}} (n3_2);
  \path (n2_2) edge [->] node[right] {\scalebox{1.3}{$\varphi_6$}} (n1_1);
  \path (n2_2) edge [->] node[above] {\scalebox{1.3}{$\varphi_5$}} (n1_2);
  \path (n1_1) edge [->] node[above] {\scalebox{1.3}{$\varphi_7$}} (n2_1);
  \path (n1_1) edge [->] node[above] {\scalebox{1.3}{$\varphi_8$}} (n3_1);

  \draw [->] (n2_1) ..controls (23.796bp,9.3918bp) and (28.588bp,5.0967bp)  .. (34.0bp,2.8922bp) .. controls (43.467bp,-0.96406bp) and (47.533bp,-0.96406bp)  .. (57.0bp,2.8922bp) .. controls (59.199bp,3.7878bp) and (61.295bp,5.0284bp)  .. (n1_1);
  \draw (45.5bp,11.892bp) node {\scalebox{1.3}{$\varphi_{6}$}};

  \draw [->] (n1_2) ..controls (23.796bp,77.392bp) and (28.588bp,73.097bp)  .. (34.0bp,70.892bp) .. controls (43.467bp,67.036bp) and (47.533bp,67.036bp)  .. (57.0bp,70.892bp) .. controls (59.199bp,71.788bp) and (61.295bp,73.028bp)  .. (n2_2);
  \draw (45.5bp,79.892bp) node {\scalebox{1.3}{$\varphi_{1}$}};

\end{tikzpicture}
}
  } 
\end{minipage}
\vspace*{-0.2cm}
\[
\begin{array}{r@{}l}
\varphi_5\equiv&  x>0 {\wedge} y<z {\wedge} y'=y+1 {\wedge} \nc{x,z} \\
\varphi_6\equiv&  x>0 {\wedge} y\geq z {\wedge} x'=x-1 {\wedge} \nc{y,z} \\
\varphi_7\equiv&  x>0 {\wedge} y\geq z {\wedge}  \nc{x,y,z} \\
\varphi_8\equiv&  x\leq 0 {\wedge} y\geq z {\wedge} \nc{x,y,z} \\
\end{array}
\]
\end{minipage}

\hrule

\begin{minipage}{0.63\textwidth}
\begin{lstlisting}
int search(int q[], int n, int h, int t, int v) {
  assert(n>0 && h<n && t<n && h>=0 && t>=0);
  while (h != t && q[t] != v) {
$\label{ex:se:inch}$    if ( h < n-1 ) h++;$\label{ex:se:else}$ else h = 0;
  }
  return q[h] == v;
}
\end{lstlisting}
\end{minipage}
~
\hspace{-0.3cm}
\begin{minipage}{0.34\textwidth}
\vspace{-0.3cm}
\hspace{-0.3cm}
  \begin{minipage}{0.305\textwidth}
    \fbox{
      \vspace{-0.3cm}
      \hspace{-0.3cm}
      \scalebox{.35}{%

\begin{tikzpicture}[>=latex,line join=bevel,]
"\Large"%
  \node (n0) at (12.0bp,140.5bp) [draw,fill=green,circle] {\scalebox{1.4}{$\node{n_0}$}};
  \node (n3) at (12.0bp,10.5bp) [draw,circle] {\scalebox{1.4}{$\node{n_3}$}};
  \node (n1) at (12.0bp,75.5bp) [draw,circle] {\scalebox{1.4}{$\node{n_1}$}};
  \node (n2) at (74.0bp,75.5bp) [draw,circle] {\scalebox{1.4}{$\node{n_2}$}};
  \node (trs) at (74.0bp,140.5bp) [] {\scalebox{1.4}{\cbox{$\trsys$}}};

  \draw [->] (n1) ..controls (30.839bp,75.5bp) and (42.726bp,75.5bp)  .. (n2);
  \draw (43.0bp,87.5bp) node {\scalebox{1.4}{$\varphi_1$}};
  \draw [->] (n1) ..controls (21.093bp,61.97bp) and (25.845bp,56.966bp)  .. (31.5bp,54.5bp) .. controls (40.87bp,50.414bp) and (45.13bp,50.414bp)  .. (54.5bp,54.5bp) .. controls (56.974bp,55.579bp) and (59.275bp,57.144bp)  .. (n2);
  \draw (43.0bp,63.5bp) node {\scalebox{1.4}{$\varphi_2$}};

  \path (n0) edge [->] node[left] {\scalebox{1.4}{$\varphi_0$}} (n1);
  \path (n1) edge [->] node[left] {\scalebox{1.4}{$\varphi_5$}} (n3);

  \draw [->] (n2) ..controls (64.907bp,89.03bp) and (60.155bp,94.03bp)  .. (54.5bp,96.5bp) .. controls (45.13bp,100.59bp) and (40.87bp,100.59bp)  .. (31.5bp,96.5bp) .. controls (29.026bp,95.42bp) and (26.725bp,93.86bp)  .. (n1);
  \draw (43.0bp,111.5bp) node {\scalebox{1.4}{$\varphi_3$}};
  \draw [->] (n2) ..controls (70.777bp,54.08bp) and (66.031bp,38.07bp)  .. (54.5bp,30.5bp) .. controls (45.955bp,24.89bp) and (40.045bp,24.89bp)  .. (31.5bp,30.5bp) .. controls (23.212bp,35.941bp) and (18.43bp,45.742bp)  .. (n1);
  \draw (43.0bp,39.5bp) node {\scalebox{1.4}{$\varphi_4$}};
\end{tikzpicture}
}
      \hspace{-0.28cm}
    }
  \end{minipage}
  ~
  \hspace{-0.3cm}
  \begin{minipage}{0.46\textwidth}
    \fbox{
      \hspace{-0.2cm}
      \scalebox{.485}{%

\begin{tikzpicture}[>=latex,line join=bevel,]
"\Large"%
  \node (n0) at (10.5bp,91.5bp) [draw,fill=green,circle] {\scalebox{1.2}{$\node{n_0}$}};
  \node (trs) at (18.5bp,12.5bp) [draw,draw=none] {\scalebox{1.2}{\cbox{$\trsys_{pe}$}}};
  \node (n1_1) at (140.5bp,12.5bp) [draw,circle] {\scalebox{1.2}{$\node{n_1^1}$}};
  \node (n2_2) at (74.5bp,12.5bp) [draw,circle] {\scalebox{1.2}{$\node{n_2^2}$}};
  \node (n2_1) at (140.5bp,91.5bp) [draw,circle] {\scalebox{1.2}{$\node{n_2^1}$}};
  \node (n1_2) at (74.5bp,91.5bp) [draw,circle] {\scalebox{1.2}{$\node{n_1^2}$}};
  \draw [->] (n2_1) ..controls (135.09bp,73.959bp) and (133.33bp,67.207bp)  .. (132.5bp,61.0bp) .. controls (131.44bp,53.07bp) and (131.44bp,50.93bp)  .. (132.5bp,43.0bp) .. controls (132.87bp,40.188bp) and (133.44bp,37.263bp)  .. (n1_1);
  \draw (144.0bp,52.0bp) node {\scalebox{1.2}{$\varphi_{3}$}};
  \draw [->] (n0) ..controls (29.395bp,91.5bp) and (41.305bp,91.5bp)  .. (n1_2);
  \draw (41.5bp,103.5bp) node {\scalebox{1.2}{$\varphi_{0}$}};
  \draw [->] (n2_2) ..controls (74.5bp,36.486bp) and (74.5bp,54.28bp)  .. (n1_2);
  \draw (86.0bp,52.0bp) node {\scalebox{1.2}{$\varphi_{6}$}};
  \draw [->] (n2_2) ..controls (95.808bp,12.5bp) and (107.38bp,12.5bp)  .. (n1_1);
  \draw (107.5bp,24.5bp) node {\scalebox{1.2}{$\varphi_{7}$}};
  \draw [->] (n1_1) ..controls (152.55bp,28.144bp) and (157.3bp,35.519bp)  .. (159.5bp,43.0bp) .. controls (161.76bp,50.675bp) and (161.76bp,53.325bp)  .. (159.5bp,61.0bp) .. controls (158.3bp,65.091bp) and (156.33bp,69.15bp)  .. (n2_1);
  \draw (172.0bp,52.0bp) node {\scalebox{1.2}{$\varphi_{2}$}};
  \draw [->] (n1_2) ..controls (53.406bp,80.912bp) and (40.413bp,72.707bp)  .. (34.5bp,61.0bp) .. controls (30.894bp,53.859bp) and (30.894bp,50.141bp)  .. (34.5bp,43.0bp) .. controls (38.704bp,34.677bp) and (46.486bp,28.124bp)  .. (n2_2);
  \draw (46.0bp,52.0bp) node {\scalebox{1.2}{$\varphi_{1}$}};
  \draw [->] (n1_2) ..controls (95.808bp,91.5bp) and (107.38bp,91.5bp)  .. (n2_1);
  \draw (107.5bp,103.5bp) node {\scalebox{1.2}{$\varphi_{2}$}};
\end{tikzpicture}
}
      \hspace{-0.4cm}
    }
  \end{minipage}
\end{minipage}
\vspace*{-1cm}

\begin{minipage}{\textwidth}
\[
\setlength\abovedisplayskip{2pt}%
\setlength\belowdisplayskip{2pt}%
\begin{array}{r@{}l}
\\
\varphi_0\equiv& h\geq 0 {\wedge} h <n  {\wedge} t\geq 0 {\wedge} t< n {\wedge} n>0 {\wedge} \nc{h,t,n}\\
\varphi_1\equiv& h>t {\wedge} \nc{h,t,n}\\
\varphi_2\equiv& h<t {\wedge} \nc{h,t,n}\\
\varphi_6\equiv& h<n-1 {\wedge} h>t {\wedge} h'=h+1 {\wedge} \nc{t,n}\\
\end{array}
\begin{array}{r@{}l}
\\
\varphi_3\equiv& h<n-1 {\wedge} h'=h+1 {\wedge} \nc{t,n}\\
\varphi_4\equiv& h\geq n-1 {\wedge} h'=0 {\wedge} \nc{t,n}\\
\varphi_5\equiv& h=t {\wedge} \nc{h,t,n}\\
\varphi_7\equiv& h=n-1 {\wedge} h'=0 {\wedge} \nc{t,n}\\
\end{array}
\]
\end{minipage}

\hrule
\smallskip

\begin{minipage}{0.49\textwidth}
\vspace{-0.3cm}
\begin{lstlisting}
void randomwalk(int n) {
  int w=1, i=1, z=n, c=0;
  while ( 1 <= w && w <= 2 ) {
$\label{ex:rw:chc}$    c=nondet(); assert(0<=c && c<=1);
$\label{ex:rw:if1}$    if ( z >= 1 ) z--;
$\label{ex:rw:if2}$    else if ( i <= 0 ) { i=2; z=n; }
$\label{ex:rw:if3}$    else if ( i >= 1 && c <= 0 ) i--;
$\label{ex:rw:if4}$    else break;
$\label{ex:rw:mdw}$    if ( c <= 0 ) w--; else  w++;
  }
}
\end{lstlisting}
\vspace{-0.55cm}
\[
\setlength\abovedisplayskip{2pt}%
\setlength\belowdisplayskip{2pt}%
\begin{array}{r@{}l}
\varphi_0\equiv&  i'=1 {\wedge} w'=1 {\wedge} z'=n {\wedge} c'=0 {\wedge} \nc{n}\\
\varphi_1\equiv& w\leq 2 {\wedge} \nc{c,i,n,w,z}\\
\varphi_2\equiv& w\geq 1 {\wedge} c'\geq 0 {\wedge} c'\leq 1 {\wedge} \nc{i,n,w,z}\\
\varphi_3\equiv& z < 1 {\wedge} \nc{c,i,n,w,z}\\
\varphi_4\equiv& z\geq 1 {\wedge} z'=z-1 {\wedge} \nc{c,i,n,w}\\
\varphi_5\equiv&  i\geq 1 {\wedge} c\leq 0 {\wedge} i'=i-1 {\wedge} \nc{c,n,w,z}\\
\end{array}
\]
\end{minipage}
~
\begin{minipage}{0.46\textwidth}
  \hspace{-0.5cm}
\begin{minipage}{0.45\textwidth}
  \fbox{
    \hspace{-0.2cm}
    \scalebox{.41}{%
\begin{tikzpicture}[>=latex,line join=bevel,]
"\Large"%
  \node (n3) at (157.5bp,12.0bp) [draw,circle] {\scalebox{1.3}{$\node{n_3}$}};
  \node (n0) at (10.5bp,177.0bp) [draw,fill=green,circle] {\scalebox{1.3}{$\node{n_0}$}};
  \node (n1) at (61.5bp,177.0bp) [draw,circle] {\scalebox{1.3}{$\node{n_1}$}};
  \node (trs) at (10.5bp,142.0bp) [draw,draw=none] {\scalebox{1.3}{\cbox{$\trsys$}}};
  \node (n2) at (157.5bp,177.0bp) [draw,circle] {\scalebox{1.3}{$\node{n_2}$}};
  \node (n5) at (61.5bp,12.0bp) [draw,circle] {\scalebox{1.3}{$\node{n_5}$}};
  \node (n4) at (109.5bp,67.0bp) [draw,circle] {\scalebox{1.3}{$\node{n_4}$}};
  \node (n6) at (109.5bp,122.0bp) [draw,circle] {\scalebox{1.3}{$\node{n_6}$}};
  \path [->] (n1) edge (n6);
  \draw (97.0bp,153.5bp) node {\scalebox{1.3}{$\varphi_{10}$}};
  \path [->] (n1) edge (n2);
  \draw (109.0bp,189.0bp) node {\scalebox{1.3}{$\varphi_{1}$}};
  \draw [->] (n4) .. controls (76.296bp,43.987bp) and (72.157bp,37.772bp)  .. (n5);
  \draw (80.0bp,60.5bp) node {\scalebox{1.3}{$\varphi_{6}$}};
  \draw [->] (n4) ..controls (115.64bp,48.542bp) and (106.4bp,34.505bp)   .. (n5);
  \draw (115.0bp,33.5bp) node {\scalebox{1.3}{$\varphi_{5}$}};
  \path [->] (n2) edge (n6);
  \draw (145.0bp,143.5bp) node {\scalebox{1.3}{$\varphi_{11}$}};
  \path [->] (n2) edge (n3);
  \draw (168.0bp,94.5bp) node {\scalebox{1.3}{$\varphi_{2}$}};
  \path [->] (n3) edge (n5);
  \draw (109.0bp,5.0bp) node {\scalebox{1.3}{$\varphi_{4}$}};
  \draw [->] (n5) ..controls (43.07bp,36.096bp) and (19.212bp,71.642bp)  .. (26.5bp,103.5bp) .. controls (31.159bp,123.87bp) and (42.313bp,145.17bp)  .. (n1);
  \draw (40.0bp,94.5bp) node {\scalebox{1.3}{$\varphi_{7}$}};
  \draw [->] (n5) ..controls (61.5bp,49.029bp) and (61.5bp,120.61bp)  .. (n1);
  \draw (75.0bp,94.5bp) node {\scalebox{1.3}{$\varphi_{8}$}};
  \path [->] (n4) edge (n6);
  \draw (121.0bp,94.5bp) node {\scalebox{1.3}{$\varphi_{9}$}};
  \draw [->] (n0) ..controls (26.86bp,177.0bp) and (33.81bp,177.0bp)  .. (n1);
  \draw (36.0bp,189.0bp) node {\scalebox{1.3}{$\varphi_{0}$}};
  \path [->] (n3) edge (n4);
  \draw (146.0bp,39.5bp) node {\scalebox{1.3}{$\varphi_{3}$}};
\end{tikzpicture}
}
    \hspace{-0.2cm}
  }
\end{minipage}
~
\begin{minipage}{0.48\textwidth}
  \fbox{
    \hspace{-0.2cm}
    \scalebox{.66}{%

\begin{tikzpicture}[>=latex,line join=bevel,]
"\Large"%
  \node (n0) at (16.0bp,95.0bp) [draw,fill=green,ellipse] { $\node{n_0}$};
  \node (box1) at (65.0bp,95.0bp) [draw,ellipse,rectangle, minimum width=1cm, minimum height = 1.25cm] { $(a)$};
  \node (box3) at (125.0bp,95.0bp) [draw,ellipse,rectangle, minimum width=1cm, minimum height = 1.25cm] { $(c)$};
  \node (n6_2) at (65.0bp,30.0bp) [draw,ellipse] {$\node{n_6^2}$};
  \node (n6_1) at (125.0bp,30.0bp) [draw,ellipse] {$\node{n_6^1}$};
  \node (trs) at (16.0bp,30.0bp) [draw,draw=none] { \cbox{$\trsys_{pe}$} } ;

  \path (n0) edge [->,dashed] (box1);
  \path (box1) edge [->,dashed] node[left] {(b)} (n6_2);
  \path (box3) edge [->,dashed]  node[left] {(d)}(n6_1);
  \path (box1) edge [->,dashed] node[above] {(b)} (box3);

  \path (box1) edge [->,dashed, loop above] (box1);

  \path (box3) edge [->,dashed, loop above] (box3);
\end{tikzpicture}
}
    \hspace{-0.2cm}
  }
\end{minipage}
\[
\begin{array}{r@{}l}
\varphi_6\equiv&  i\leq 0 {\wedge} i'=2 {\wedge} z'=n {\wedge} \nc{c,n,w}\\
\varphi_7\equiv&  c\leq 0 {\wedge} w'=w-1 {\wedge} \nc{c,i,n,z}\\
\varphi_8\equiv&  c> 0 {\wedge} w'=w+1 {\wedge} \nc{c,i,n,z}\\
\varphi_9\equiv&  i\geq 1 {\wedge} c> 0 {\wedge} \nc{c,i,n,w,z}\\
\varphi_{10}\equiv& w > 2 {\wedge} \nc{c,i,n,w,z}\\
\varphi_{11}\equiv& w < 1 {\wedge} \nc{c,i,n,w,z}\\
\end{array}
\]
\end{minipage}

\hrule
\smallskip

\begin{minipage}{0.46\textwidth}
\vspace{-1.5cm}
\begin{lstlisting}
void phases2(int x, int y, int z) {    
  while( x >= 0 ) {
    x = x + y;
    y = y + z;
    z = z - 1;
  }
}
\end{lstlisting}
\end{minipage}
\begin{minipage}{0.52\textwidth}
  \vspace{-0.09cm}
  \hspace{-0.25cm}
  \begin{minipage}{0.2\textwidth}
    \fbox{
      \hspace{-0.38cm}
      \scalebox{.68}{%

\begin{tikzpicture}[>=latex,line join=bevel,]
"\Large"%
  \node (n0) at (10.0bp,137.0bp) [draw,fill=green,circle] {$\node{n_0}$};
  \node (n1) at (10.0bp,77.0bp) [draw,circle] {$\node{n_1}$};
  \node (n2) at (10.0bp,17.0bp) [draw,circle] {$\node{n_2}$};
  \node (trs) at (-18.0bp,137.0bp) [draw,draw=none] {\cbox{$\trsys$}};

  \path (n1) edge [->,loop left] (n1); %
  \draw (-14.0bp,94.0bp) node {$\varphi_1$};
  \path (n1) edge [->] node[left] {$\varphi_2$} (n2);
  \path (n0) edge [->] node[left] {$\varphi_0$} (n1);
\end{tikzpicture}
}
      \hspace{-0.27cm}
    }
  \end{minipage}
  \begin{minipage}{0.26\textwidth}
    \fbox{
      \hspace{-0.29cm}
      \scalebox{.41}{%

\begin{tikzpicture}[>=latex,line join=bevel,]
"\huge"%
  \node (n0) at (95.0bp,157.5bp) [draw,fill=green,circle] {$\node{n_0}$};
  \node (n1a) at (95.0bp,90.5bp) [draw,circle] {$\node{n_{1a}}$};
  \node (trs) at (52.0bp,157.5bp) [draw,draw=none] {\cbox{$\trsys'$}};
  \node (n1) at (95.0bp,12.0bp) [draw,circle] {$\node{n_{1}}$};
  \node (n2) at (95.0bp,-50.0bp) [draw,circle] {$\node{n_2}$};
  \draw [->] (n0) ..controls (95.0bp,140.0bp) and (95.0bp,127.0bp)  .. (n1a);
  \draw (110.5bp,125.0bp) node {\scalebox{0.93}{$\varphi_0$}};
  \draw [->] (n1a) ..controls (72.0bp,81.0bp) and (57.0bp,72.0bp)  .. (50.0bp,60.0bp) .. controls (46.0bp,52.0bp) and (46.0bp,49.0bp)  .. (50.0bp,42.0bp) .. controls (55.0bp,32.0bp) and (65.0bp,25.0bp)  .. (n1);
  \draw (58.56bp,83.0bp) node {\scalebox{0.93}{$\varphi_3$}};

  \draw [->] (n1a) ..controls (81.0bp,76.0bp) and (74.0bp,68.0bp)  .. (71.0bp,60.0bp) .. controls (69.0bp,52.0bp) and (69.0bp,49.0bp)  .. (71.0bp,42.0bp) .. controls (73.0bp,37.0bp) and (76.0bp,32.0bp)  .. (n1);
  \draw (58.56bp,51.0bp) node {\scalebox{0.93}{$\varphi_4$}};
  \draw [->] (n1a) ..controls (109.0bp,76.0bp) and (116.0bp,68.0bp) .. (119.0bp,60.0bp) .. controls (121.0bp,52.0bp) and (121.0bp,49.0bp)  .. (119.0bp,42.0bp) .. controls (117.0bp,37.0bp) and (114.0bp,32.0bp)  .. (n1);
  \draw (132.56bp,51.0bp) node {\scalebox{0.93}{$\varphi_6$}};
  \draw [->] (n1a) ..controls  (122.0bp,77.0bp) and (159.0bp,63.0bp) .. (145.0bp,42.0bp) .. controls (138.0bp,31.0bp) and (117.0bp,20.0bp) .. (n1);
  \draw (132.56bp,83.0bp) node {\scalebox{0.93}{$\varphi_5$}};
\path (n1) edge [->] (n1a);
\draw (105.0bp,52.0bp) node {\scalebox{0.93}{$\varphi_1$}};
\path (n1) edge [->] node[left] {\scalebox{0.93}{$\varphi_2$}}(n2);
\end{tikzpicture}
}
      \hspace{-0.4cm}
    }
  \end{minipage}
  \begin{minipage}{0.28\textwidth}
    \fbox{
      \hspace{-0.2cm}
      \scalebox{.5}{%

\begin{tikzpicture}[>=latex,line join=bevel,]
"\Large"%
  \node (n1_4) at (95.0bp,182.53bp) [draw,circle] {\scalebox{1.1}{$\node{n_{1a}^4}$}};
  \node (n1_3) at (14.0bp,100.53bp) [draw,circle] {\scalebox{1.1}{$\node{n_{1a}^3}$}};
  \node (trs) at (176.0bp,142.53bp) [draw,draw=none] {\scalebox{1.3}{\cbox{$\trsys'_{pe}$}}};
  \node (n0) at (14.0bp,182.53bp) [draw,fill=green,circle] {\scalebox{1.1}{$\node{n_0}$}};
  \node (n2_2) at (14.0bp,18.531bp) [draw,circle] {\scalebox{1.1}{$\node{n_2^2}$}};
  \node (n1_1) at (95.0bp,18.531bp) [draw,circle] {\scalebox{1.1}{$\node{n_{1a}^1}$}};
  \node (n2_1) at (176.0bp,18.531bp) [draw,circle] {\scalebox{1.1}{$\node{n_2^1}$}};
  \node (n2_3) at (176.0bp,182.53bp) [draw,circle] {\scalebox{1.1}{$\node{n_2^3}$}};
  \node (n1_2) at (95.0bp,100.53bp) [draw,circle] {\scalebox{1.1}{$\node{n_{1a}^2}$}};

  \path (n1_4) edge [->] node[right] {\scalebox{1.3}{$\varphi_3$}}(n1_3);
  \path (n1_4) edge [->] node[right] {\scalebox{1.3}{$\varphi_4$}}(n1_2);
  \path (n1_2) edge [->] node[right] {\scalebox{1.3}{$\varphi_5$}}(n1_1);
  
  \draw [->] (n1_4) ..controls (144.05bp,153.9bp) and (167.23bp,118.01bp)  .. (158.0bp,86.531bp) .. controls (152.63bp,68.212bp) and (140.53bp,50.157bp)  .. (n1_1);
  \draw (172.5bp,100.53bp) node {\scalebox{1.3}{$\varphi_5$}};

  \path (n1_1) edge [->,loop left]  (n1_1);
  \draw (53.0bp,34.0bp) node {\scalebox{1.3}{$\varphi_5$}};
  \path (n1_3) edge [->] node[above] {\scalebox{1.3}{$\varphi_4$}}(n1_2);
  \path (n0) edge [->] node[above] {\scalebox{1.3}{$\varphi_0$}}(n1_4);

  \path (n1_2) edge [->,loop right] node[above] {\scalebox{1.3}{$\varphi_4$}} (n1_2);

  \path (n1_4) edge [->] node[above] {\scalebox{1.3}{$\varphi_2$}}(n2_3);
  \path (n1_3) edge [->] node[right] {\scalebox{1.3}{$\varphi_2$}}(n2_2);
  \path (n1_1) edge [->] node[above] {\scalebox{1.3}{$\varphi_6$}}(n2_1);

  \path (n1_3) edge [->,loop above] node[above] {\scalebox{1.3}{$\varphi_3$}} (n1_3);
  \path (n1_3) edge [->] node[above] {\scalebox{1.3}{$\varphi_5$}}(n1_1);

\end{tikzpicture}
}
      \hspace{-0.25cm}
    }
  \end{minipage}
\end{minipage}

\begin{minipage}{0.52\textwidth}
\vspace{-1.7cm}
\[
\begin{array}{r@{}l}
\varphi_0\equiv&  \nc{x,y,z} \\
\varphi_2\equiv&  x<0 {\wedge} \nc{x,y,z} \\
\varphi_3\equiv&  z\geq0 {\wedge} \nc{x,y,z} \\
\varphi_4\equiv&  z<0 {\wedge} y\geq 0 {\wedge} \nc{x,y,z} \\
\varphi_1\equiv&  x\geq0 {\wedge} x'=x+y {\wedge} y'=y+z {\wedge} z'=z-1\\
\end{array}
\]
\end{minipage}
~
\begin{minipage}{0.42\textwidth}
\vspace{-0.2cm}
\[
\setlength\abovedisplayskip{2pt}%
\setlength\belowdisplayskip{2pt}%
\begin{array}{r@{}l}
\varphi_5\equiv&  z<0 {\wedge} y<0 {\wedge} x\geq 0 {\wedge} \nc{x,y,z} \\
\varphi_6\equiv&  z<0 {\wedge} y<0 {\wedge} x<0 {\wedge} \nc{x,y,z} \\
\end{array}
\]
\end{minipage}

\hrule

\caption{Example programs. (a) \lst{phases1}: a loop with $2$ phases; (b) \lst{search}: search in a circular queue; (c) \lst{randomwalk}: a random walk simulation; (d) \lst{phases2}: a loop with $3$ phases.}
\label{fig:progs}

\end{figure}
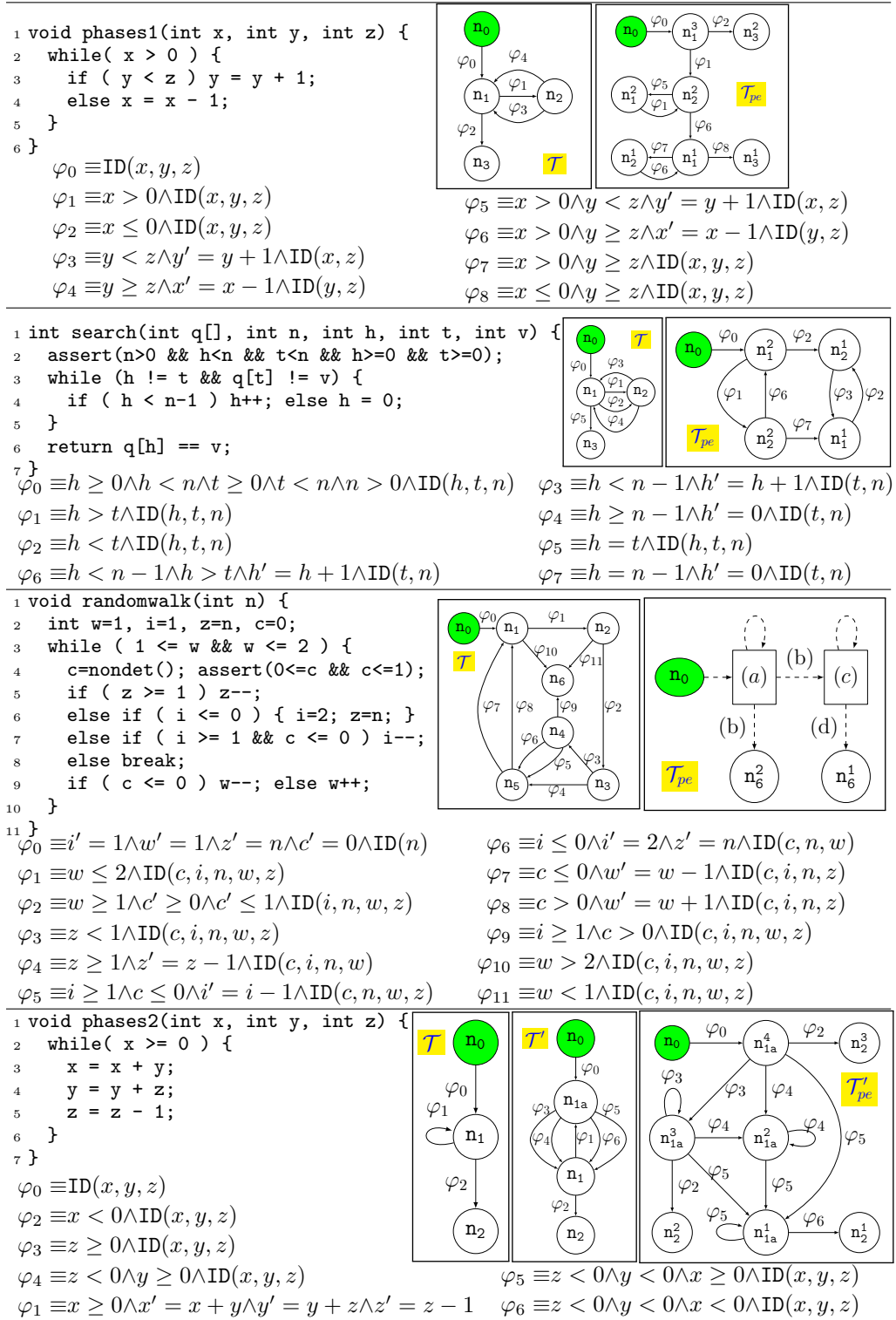

\section{Preliminaries}
\label{sec:prelim}

In this section we define the programming model, integer transition
systems, that we will use throughout this paper and provide some
necessary related background.

We let $V$ be a fixed set of integer-valued variables
$\{x_1,\ldots,x_n\}$.
A linear constraint $\psi$ is of the form
$a_0+a_1x_1+\cdots+a_nx_n \diamond 0$
where $\diamond \in \{>,<,\ge,\le,=\}$, $a_i$ are integer
coefficients, and $x_i \in V$.
An assignment $\sigma:V \mapsto \Z$ is a mapping that assigns integer
values to variables. We say that $\sigma$ is a solution for $\psi$ if
$a_0+\sum_ia_i\sigma(x_i) \diamond 0$ is $\mathit{true}$. The set of
all solutions for $\psi$ is denoted by $\sols{\psi}$.
A \emph{formula} $\varphi$ is a conjunction of linear constraints of the form
$\psi_1\wedge\cdots\wedge \psi_k$, and its set of solutions is
$\sols{\varphi}=\sols{\psi_1}\cap\cdots\cap\sols{\psi_k}$.
We write $\varphi_1 \models \varphi_2$ for
$\sols{\varphi_1} \subseteq \sols{\varphi_2}$.
The projection of a formula $\varphi$ onto variables $X \subseteq V$,
denoted as $\proj{X}{\varphi}$, is a formula $\varphi'$ such that
$\sols{\varphi'}$ coincide with
$\{ \sigma|_{X} \mid \sigma \in \sols{\varphi} \}$ where $\sigma|_{X}$
restricts the domain of $\sigma$ to $X$.
For example, restricting
$x_1 \le x_2 \wedge x_2 \le 100 \wedge x_3 = x_2+10$ to $\{x_1,x_3\}$
results in $x_1 \le 100 \wedge x_3 \le 110$.
Note that $\varphi'$ can be effectively computed using off-the-shelf
libraries for manipulating linear constraints~\cite{BagnaraHZ08}.
We use $\varphi[x_i/x_j]$ to denote the formula obtained from
$\varphi$ by renaming variable $x_i$ to $x_j$.

An integer transition system $\trsys$ (\its for short) is a
control-flow graph where edges are labeled with formulas.
Formally, $\trsys = \tuple{V,N,\node{n_0},E}$ where $V$ is a set of
program variables, $N$ is a set of nodes, $\node{n_0}\in N$ is the
entry node, and $E$ is a set of labeled edges. An
edge is of the form $\edge{\node{n_i}}{\varphi}{\node{n_j}}$ such that
$\node{n_i},\node{n_j}\in N$ and $\varphi$ is \emph{formula} over
variables $V \cup V'$ where $V'=\{x_i' \mid x_i \in V\}$. 
We assume that the entry node $\node{n_0}$ has no incoming
edges.
We write $\nc{x_1,\ldots,x_n}$ for
$x_1=x_1'\wedge\ldots\wedge x_n=x_n'$, to indicate that variables
$x_1,\ldots,x_n$ do not change their values in a given edge.

A program state is a pair $(\node{n_i},\sigma)$, where
$\node{n_i}\in N$ and $\sigma:V \mapsto\Z$ is an assignment.
There is a transition (i.e. a valid execution step) from
$s_1=(\node{n_i},\sigma)$ to $s_2=(\node{n_j},\sigma')$, denoted as
$s_1\rightarrow s_2$, if there is an edge
$\edge{\node{n_i}}{\varphi}{\node{n_j}} \in E$ such that
$\wedge_{x_i\in V}(x_i=\sigma(x_i)\wedge x_i'=\sigma'(x_i)) \models
\varphi$.
The primed
variables $V'$ in $\varphi$ refer to the program state following the
transition. 

A trace is a sequence of states
$s_0\rightarrow s_1 \rightarrow \cdots$ such that
$s_i \rightarrow s_{i+1}$ is a transition. We write
$s_i \rightarrow^* s_j$ to indicate that state $s_j$ is reachable from
state $s_i$.
An invariant (wrt. an initial state $s_0$) for a given node
$\node{n_i}\in N$ is a formula $\varphi$ over the program variables,
and is guaranteed to holds whenever the execution reaches
$\node{n_i}$, i.e., if $s_0 \rightarrow^* (\node{n_i},\sigma)$ then
$\sigma \in \sols{\varphi}$.
An \its is \emph{terminating} if there are no traces of infinite
length starting in $(\node{n_0},\sigma)$.

The complexity (also called cost) of an \its is typically defined in terms
of the length of its traces. In order to model cost in a way that is
amenable to program transformations, we can add an extra special variable
$x_{n+1}$ whose value is $0$ in the initial state and which is
incremented by $1$ in every transition ($x_{n+1}$ can be modified in
other ways to capture different cost models).
A function $f:\Z^n\mapsto \R_+$ is an upper-bound on the cost of
\trsys if for any trace
$(\node{n_0},\sigma) \rightarrow^* (\node{n},\sigma')$, we have
$f(\sigma(x_1),\ldots,\sigma(x_n)) \ge \sigma'(x_{n+1})$.

\section{Partial Evaluation of Integer Transition Systems}
\label{sec:pe}

In this section we give a brief description of the partial evaluation technique
(for Horn clauses)  of~\citeN{Gallagher19}, and discuss its
use for \cfr of \itss.
A constrained Horn clause (\chc) has the form
$q_0(\bar{x}_0) \leftarrow \varphi,
q_1(\bar{x}_1),\ldots,q_k(\bar{x}_k)$, where $q_i$ are predicate names
(all of arity $n$ for simplicity) and $\varphi$ is a constraint. 
Here we assume that the $\bar{x}_i$ are tuples of
integer variables, and $\varphi$ is a formula over these variables, where 
the formula is a conjunction of linear constraints as defined in Sect.~\ref{sec:prelim}.
We call $q_0(\bar{x}_0)$ the head, and
``$\varphi,q_1(\bar{x}_1),\ldots,q_k(\bar{x}_k)$'' the body. 
We say that the set of \chcs with head predicate $q_0$ defines $q_0$. %
A \chc program is a set of \chcs such that one predicate is marked as
the entry. The call graph induced by a \chc program is a directed
graph whose nodes are the predicate names, and there is an edge from
$q_i$ to $q_j$ if $q_i$ is defined by a \chc including $q_j$ in its
body.
We say that $q_i$ is a \emph{loop head} if it has a \emph{back edge}
in the corresponding call-graph wrt. depth-first traversal from the
entry predicate.

The technique of~\citeN{Gallagher19} takes as input a \chc program and
a constraint on the entry predicate, and returns a partially evaluated \chc program that
preserves the computational behavior of the original program with the
given entry.  In particular, it preserves termination and complexity of the
original.
It is an online partial evaluation algorithm, meaning that control decisions are 
made on the fly, and it yields a \emph{polyvariant} partially evaluated program, 
meaning that a single predicate $q_i$ from the input program can 
result in several ``versions" of $q_i$ in the output program.

It is not possible to create versions for all
reachable contexts as there may be infinitely many.
During partial evaluation, when reaching a call
$q(\bar{x})$ with a context $\varphi$ (i.e. constraints on variables
$\bar{x}$) the algorithm decides whether to create a version of $q(\bar{x})$ for
this context, and replace the corresponding call by one to
this version, or to \emph{abstract} the context. 

\citeN{Gallagher19} suggests the use of property-based abstraction in
order to guarantee the generation of a finite number of versions.
For a predicate $q(\bar{x})$ it assigns beforehand a finite set of
properties $\{\varphi_1,\ldots,\varphi_k\}$, where each $\varphi_i$ is
a formula over $\bar{x}$, and when reaching $q(\bar{x})$ with a context $\varphi$,
instead of creating a version for $\varphi$ it creates one for
$\alpha(\varphi)=\wedge\{ \varphi_i \mid \varphi \models \varphi_i
\}$. This guarantees that there are at most $2^k$ versions for
$q(\bar{x})$.
For example, if predicate $p(x,y,z)$ is assigned properties $x>0$ and
$y \ge z$, it would then have up to $4$ versions corresponding to
(abstract) contexts $x>0$, $y \ge z$, $x>0 \wedge y \ge z$ and
$\mathit{true}$.
In practice, it is enough to apply property-based abstraction to loop
head predicates as they cut all cycles; other predicates can be
specialized wrt. to all contexts that are encountered during the
evaluation. In addition, partial evaluation might unfold deterministic sequences of
calls into a single \chc.

An \its $\trsys$ can be translated into a \chc program
$\chct{\trsys}$ by translating each edge
$\edge{\node{n_i}}{\varphi}{\node{n_j}}$ into a \chc
$q_{\node{n_i}}(\bar{x}) \leftarrow \varphi,
q_{\node{n_j}}(\bar{x}')$, and marking $q_{\node{n_0}}$ as the entry.
Note that the control-flow graph of $\trsys$ is equivalent to the
call-graph of $\chct{\trsys}$.
Observe that $\chct{\trsys}$ is linear, that is, each clause has only one
call in the body. Linearity guarantees that the specialized version is
also linear, and thus can be converted back into an \its $\trsys_{pe}$
in a similar way.
Soundness of partial evaluation guarantees equivalence between the
traces of $\trsys$ and $\trsys_{pe}$, in particular, there is a
infinite trace in $\trsys$ iff there is one in $\trsys_{pe}$. The
complexity is preserved as well (when modeled as discussed in
Sect.~\ref{sec:prelim}).

\begin{example}
\label{ex:pe}
We now describe the partial evaluation algorithm for Horn clauses 
using a worked example,
assuming that the clauses are linear, as they are generated from an \its as just described.

Consider the method \lst{phases1} of Fig.~\ref{fig:progs}(a) and its
corresponding \its $\trsys$. As discussed in
Sect.~\ref{sec:intro}, the execution of this loop has two phases.
Our aim is
to use partial evaluation to split the
loop into two explicit phases (see the corresponding $\trsys_{pe}$ of
Fig.~\ref{fig:progs}(a)).
Translating this \its into a \chc program results in:%
\[
\setlength\abovedisplayskip{2pt}%
\setlength\belowdisplayskip{2pt}%
\begin{array}{llllll}
q_{\node{n_0}}(\bar{x}) \leftarrow \varphi_0, q_{\node{n_1}}(\bar{x}'). &~~
q_{\node{n_1}}(\bar{x}) \leftarrow \varphi_1, q_{\node{n_2}}(\bar{x}'). &~~
q_{\node{n_1}}(\bar{x}) \leftarrow \varphi_2, q_{\node{n_3}}(\bar{x}'). \\ 
q_{\node{n_2}}(\bar{x}) \leftarrow \varphi_3, q_{\node{n_1}}(\bar{x}'). &~~
q_{\node{n_2}}(\bar{x}) \leftarrow \varphi_4, q_{\node{n_1}}(\bar{x}'). \\
\end{array}
\]
where $\bar{x}=\tuple{x,y,z}$, $\bar{x}'=\tuple{x',y',z'}$, and
each $\varphi_i$ is as in Fig.~\ref{fig:progs}(a).
Note that the only loop head predicate is $q_{\node{n_1}}$
($\node{n_1}$ has a back edge when traversing in $\trsys$ starting at
$\node{n_0}$).
Let us fix the properties of $q_{\node{n_1}}(x,y,z)$ to be $\Psi = 
\{x>0, y < z, y \ge z\}$, which are the properties of the variables at loop entry
that determine which loop path or loop exit is taken.

The partial evaluation algorithm performs a sequence of iterations. 
The input to each iteration is a set of ``versions" of predicates, where each version
is a pair $\langle q(\bar{x}), \varphi \rangle$, where $q$ is a program predicate and
$\varphi$ is a constraint on its arguments. Each iteration performs an
\emph{unfolding} for each version, followed by a property-based \emph{abstraction} 
of the resulting
calls using $\Psi$, yielding a new set of versions that is input to the 
following iteration. The iterations
end when no new versions are produced by an unfolding.
The unfold operation for 
a version $\langle q(\bar{x}), \varphi \rangle$ consists of expanding
calls in the body of clauses with head $q(\bar{x})$, so long as the call is deterministic 
and not a loop head, and the constraint $\varphi$ is satisfied.  

Let the initial version be $\langle q_{\node{n_0}}(x,y,z), \mathit{true} \rangle$. 
Then the iterations of the algorithm yield the following sequence of new versions.
\begin{compactenum}
\item
$\{\langle q_{\node{n_1}}(x,y,z), \mathit{true} \rangle \}$.
\item
$\{\langle q_{\node{n_2}}(x,y,z), x > 0 \rangle, \langle q_{\node{n_3}}(x,y,z), x \le 0  \rangle\}$.
\item
$\{\langle q_{\node{n_1}}(x,y,z), x > 0 \rangle, \langle q_{\node{n_1}}(x,y,z), y \ge z \rangle\}$.
\item
$\{\langle q_{\node{n_2}}(x,y,z), x > 0 \wedge y \ge z  \rangle, \langle q_{\node{n_3}}(x,y,z), x \le 0, y \ge z  \rangle\}\}$.
\item
$\emptyset$.
\end{compactenum}
We look in detail at iteration 3;  the versions $\langle q_{\node{n_2}}(x,y,z), \mathit{x > 0} \rangle$ and
$ \langle q_{\node{n_3}}(x,y,z), x \le 0 \rangle$ resulting from iteration 2 are
unfolded.  The latter yields no clauses since there is no clause with head predicate $q_{\node{n_3}}$.
The version $\langle q_{\node{n_2}}(x,y,z), \mathit{x > 0} \rangle$ is unfolded giving the clauses 
\[
\setlength\abovedisplayskip{2pt}%
\setlength\belowdisplayskip{2pt}%
\begin{array}{l}
q_{\node{n_2}}(x,y,z)  \leftarrow x>0,z>y,y'=y-1,q_{\node{n_1}}(x,y',z)\\
q_{\node{n_2}}(x,y,z)  \leftarrow x>0,y\ge z,x'=x-1,q_{\node{n_1}}(x',y,z).
\end{array}
\]
The constraints on the body calls are then collected.  The constraints on $q_{\node{n_1}}(x,y',z)$,
projected onto $\{x,y',z\}$, are $x>0$.  The constraints on $q_{\node{n_1}}(x',y,z)$, projected
onto $\{x',y,z\}$, are $x' > -1, y\ge z$. Using property-based abstraction 
with $\Psi$, the set of properties for $q_{\node{n_1}}$, we obtain (after renaming $x'$ to $x$)
\[
\setlength\abovedisplayskip{2pt}%
\setlength\belowdisplayskip{2pt}%
\begin{array}{lcl}
\alpha(x>0) = x>0 &~~~~~~~~~~& \alpha(x > -1 \wedge y\ge z) = y\ge z.
\end{array}
\]
Note that the constraint $x > -1 \wedge y\ge z$ entails $y\ge z$ but no other member of $\Psi$
and hence the constraint $x > -1$ is
abstracted away.
This yields the two versions $\langle q_{\node{n_1}}(x,y,z), x > 0 \rangle$ and 
$\langle q_{\node{n_1}}(x,y,z), y \ge z \rangle$ that are shown as the result of iteration 3.
Note that without abstraction, an infinite number of iterations could
result.  In this example, versions 
$\langle q_{\node{n_1}}(x,y,z), x > -1, y \ge z \rangle, \langle q_{\node{n_1}}(x,y,z), x > -2, y \ge z \rangle, \ldots$
would be generated.

For each version, a new predicate is generated.  
We use predicate names that carry information on the different
versions: $q_{\node{n_i^j}}$ is the $j$th version of predicate
$q_{\node{n_i}}$.The entry predicate $q_{\node{n_0}}$ is not renamed. 
For our example a suitable renaming is as follows.
\[
\setlength\abovedisplayskip{2pt}%
\setlength\belowdisplayskip{2pt}%
\begin{array}{lll|lll}
\langle q_{\node{n_1}}(x,y,z), \mathit{true} \rangle &\Rightarrow& q_{\node{n_1^3}}(x,y,z)&~~
\langle q_{\node{n_2}}(x,y,z), x > 0 \rangle  &\Rightarrow& q_{\node{n_2^2}}(x,y,z)\\
\langle q_{\node{n_1}}(x,y,z), x > 0 \rangle  &\Rightarrow& q_{\node{n_1^2}}(x,y,z) ~~& ~~
\langle q_{\node{n_2}}(x,y,z), x > 0 \wedge y \ge z  \rangle  &\Rightarrow& q_{\node{n_1^2}}(x,y,z)\\
\langle q_{\node{n_1}}(x,y,z), y \ge z \rangle  &\Rightarrow& q_{\node{n_1^1}}(x,y,z)&~~
\langle q_{\node{n_3}}(x,y,z), x \le 0\rangle  &\Rightarrow& q_{\node{n_3^2}}(x,y,z)\\
&&&~~
\langle q_{\node{n_3}}(x,y,z), x \le 0, y \ge z \rangle  &\Rightarrow& q_{\node{n_3^1}}(x,y,z)
\end{array}
\]
The head and body calls of the unfolded clauses 
for each version are then renamed, yielding the following set of clauses as the result of partial evaluation ($\varphi_i$ are as in Fig.~\ref{fig:progs}(a)):
\[
\setlength\abovedisplayskip{2pt}%
\setlength\belowdisplayskip{2pt}%
\begin{array}{llllll}
q_{\node{n_0}}(\bar{x}) \leftarrow    \varphi_0, q_{\node{n_1^3}}(\bar{x}).&~~
q_{\node{n_1^3}}(\bar{x}) \leftarrow  \varphi_2, q_{\node{n_3^2}}(\bar{x}).&~~
q_{\node{n_1^3}}(\bar{x}) \leftarrow  \varphi_1, q_{\node{n_2^2}}(\bar{x}).\\
q_{\node{n_2^2}}(\bar{x}) \leftarrow  \varphi_5, q_{\node{n_1^2}}(\bar{x}).&~~
q_{\node{n_2^2}}(\bar{x}) \leftarrow  \varphi_6, q_{\node{n_1^1}}(\bar{x}).&~~
q_{\node{n_1^2}}(\bar{x}) \leftarrow  \varphi_1, q_{\node{n_2^2}}(\bar{x}).\\
q_{\node{n_1^1}}(\bar{x}) \leftarrow  \varphi_8, q_{\node{n_3^1}}(\bar{x}).&~~
q_{\node{n_1^1}}(\bar{x}) \leftarrow  \varphi_7, q_{\node{n_2^1}}(\bar{x}).&~~
q_{\node{n_2^1}}(\bar{x}) \leftarrow  \varphi_6, q_{\node{n_1^1}}(\bar{x}).\\
\end{array}
\]
Translating these \chcs back to an \its results in $\trsys_{pe}$,
depicted in Fig.~\ref{fig:progs}(a).
\ebox
\end{example}

\paragraph{Choice of properties.} In the above example, the properties were chosen manually to give the appropriate versions
and achieve \cfr. While any choice of properties gives a sounds partial evaluation,  heuristics are needed to infer 
appropriate properties automatically.
We observe that properties relevant to \cfr are closely related to the conditions in the loop head and within the loop body. 
This leads to the following heuristics that extract
properties of $q_{\node{n_i}}$ from the constraints of
incoming and outgoing edges of node $\node{n_i}$: 
\[
\setlength\abovedisplayskip{2pt}%
\setlength\belowdisplayskip{2pt}%
\begin{array}{ll@{}l}
  \propsh(q_{\node{n_i}}) &= &~ \{ \psi \mid q_{\node{n_i}}(\bar{x}) \leftarrow \varphi,q_{\node{n_j}}(\bar{x}') \in \chct{\trsys}, \psi=\proj{\bar{x}}{\varphi} \} \\
  \propshv(q_{\node{n_i}}) &= &~ \{ x_\ell \diamond c \mid q_{\node{n_i}}(\bar{x}) \leftarrow \varphi,q_{\node{n_j}}(\bar{x}') \in \chct{\trsys}, \ell \in [1..n], \diamond \in\{\le,\ge\}, \varphi \models x_\ell \diamond c \} \\  
  \propsc(q_{\node{n_i}}) &= &~ \{ \psi[\bar{x}'/\bar{x}] \mid q_{\node{n_j}}(\bar{x}) \leftarrow \varphi,q_{\node{n_i}}(\bar{x}') \in \chct{\trsys}, \psi=\proj{\bar{x}'}{\varphi} \} \\
 \propscv(q_{\node{n_i}}) &= &~ \{ x_\ell \diamond c \mid q_{\node{n_j}}(\bar{x}) \leftarrow \varphi,q_{\node{n_i}}(\bar{x}') \in \chct{\trsys}, \ell \in [1..n], \diamond \in\{\le,\ge\},  \varphi \models x'_\ell \diamond c \} \\
\end{array}
\]
$\propsh$ infers properties by extracting conditions from the \chcs
defining $q_{\node{n_i}}$ (i.e., outgoing edge of $\node{n_i}$); this
is done by projecting the corresponding constraints on $\bar{x}$. Note
that this also captures implicit conditions that do not
appear syntactically in $\varphi$.
$\propshv$ infers properties by extracting bounds, for each variable
$x_i$, from the \chcs defining $q_{\node{n_i}}$, inferring the
minimal/maximal value for $x_i$ is done in practice using linear
programming. $\propsc$ and $\propscv$ are similar but use calls to
$q_{\node{n_i}}$ (i.e., incoming edge of $\node{n_i}$).

\begin{example}
\label{ex:prop:1}
Let us compute properties for $\node{q_{n_1}}$. The corresponding node
$\node{n_1}$ has outgoing edges labeled with $\varphi_1$ and
$\varphi_2$ and incoming edges labeled with $\varphi_0$, $\varphi_3$
and $\varphi_4$.
$\propsh(q_{\node{n_1}}) = \{\proj{\bar{x}}{\varphi_1},
\proj{\bar{x}}{\varphi_2}\} = \{ x \geq 1, x\leq 0 \}$, which is also
what we get for $\propshv$ in this case.
$\propsc(q_{\node{n_1}}) = \{\proj{\bar{x}'}{\varphi_0},
\proj{\bar{x}'}{\varphi_3}, \proj{\bar{x}'}{\varphi_4}
\}[\bar{x}'/\bar{x}] = \{ y\leq z, y\geq z \}$ and
$\propscv = \emptyset$.
\ebox
\end{example}

The above heuristics suffice for many cases in practice, but they
may be inadequate when conditions in a loop body are not directly implied by
the formulas of incoming/outgoing edges of the loop head.
E.g., assume method \lst{phase1} %
has an additional statement \lstinline{w=w+1} at the end of the loop body.
In this case, the outgoing edges of $\node{n_2}$ will not go directly
to $\node{n_1}$, but  to a new node 
connected 
 to $\node{n_1}$ with the formula
$w'=w+1 \wedge \nc{x,y,z}$. Thus, $y<z$ and $z \ge y$ are not implied
by the constraints of incoming/outgoing edges of $\node{n_1}$.

To overcome this limitation, we developed a new heuristic
$\propsdh$ that propagates all constraints in the loop bodies
backwards
to loop heads as follows:
(1) we remove all back-edges (in \chct{\trsys}), that is,
we replace each
``$q_{\node{n_j}}(\bar{x}) \leftarrow \varphi,q_{\node{n_i}}(\bar{x}')
\in \chct{\trsys}$'' by
``$q_{\node{n_j}}(\bar{x}) \leftarrow \varphi$'' if $q_{\node{n_i}}$
is a loop head, which results in a recursion-free \chc program;
(2) we compute the set of answers (the minimal model) of the resulting
\chc program and take them as properties, i.e., if
$\varphi_1,\ldots,\varphi_l$ are the answers for
$q_{\node{n_i}}(\bar{x})$, we define
$\propsdh(q_{\node{n_i}})=\{\varphi_1,\ldots,\varphi_l\}$.
For the modification of the program of
Fig.~\ref{fig:progs}(a) just described, we would get
$\propsdh = \{y<z, y\geq z, x>0\}$.
We could propagate conditions in the loop bodies %
forwards as well;
however, we do not do it since in practice we add invariants to all
transitions (which has the effect of propagating conditions forward)
before inferring properties.

In the rest of this article, we assume that we have a procedure
$\proccfr$ that receives an \its \trsys and applies \cfr via partial
evaluation as follows:
\begin{inparaenum}[\upshape(1\upshape)]
\item optionally, it computes invariants for all nodes, and adds
  them to their outgoing edges, which makes more properties visible in
  the incoming edges of nodes as discussed above;
\item generates the \chc program $\chct{\trsys}$;
\item computes properties for each loop head;
\item applies partial evaluation as described above;
\item translates the specialized \chcs into a new \its;
\item optionally, computes invariants again in order to eliminate
  unreachable nodes in the new \its; and
\item returns the new \its.
\end{inparaenum}
For simplicity, we assume that the choice of properties is provides
via global configuration, and not received as a parameter.
We might also call \proccfr with a list of nodes $N$ as a second
argument to indicate that properties should be assigned to a loop head
$q_{\node{n_i}}$ only if $\node{n_i}\in N$, this is used to apply \cfr
only to a subset of the \its.
Finally, procedure \pe can be applied iteratively on its
output, which might achieve further refinements.

\section{Application to Termination and Cost Analysis}
\label{sec:termin}

In this section we discuss how \cfr via partial evaluation
can improve the precision of
termination and cost analysis of \itss.
After a high-level description of a typical termination analyzer (based on \rankfinder~\cite{rankfinder}), we present an algorithm for termination analysis 
incorporating the \cfr procedure of Sect.~\ref{sec:pe}, and finally
discuss 
some
representative examples.
At the end of the section we discuss cost
analysis.

A termination analyzer receives an input \its $\trsys$, and separately proves
termination of each its strongly connected components (\sccs) 
using a procedure that we call \procterminscc.
Proving termination of an \scc $S$ is done by inferring
a \emph{ranking function}
$f_{\node{n_i}}:\Z^n\mapsto D$ for each node
$\node{n_i}$
mapping program states into a set $D$ ordered by a well-founded relation $>$,  such that if
$\edge{\node{n_i}}{\varphi}{\node{n_j}}$ is an edge of $S$ then
$\varphi \models f_{\node{n_i}}(\bar{x}) > f_{\node{n_j}}(\bar{x}')$.
Well-foundedness of $>$ guarantees the absence of infinite traces. 
In practice, we require
$f_{\node{n_i}}(\bar{x}) > f_{\node{n_j}}(\bar{x}')$ to hold only on a
subset of the edges that break all cycles; for other edges it is
enough that
$f_{\node{n_i}}(\bar{x}) \ge f_{\node{n_j}}(\bar{x}')$.
Even if \procterminscc fails to prove termination of $S$, 
it might prove that some of its edges cannot be taken infinitely,
and thus that any infinite trace has a suffix that uses only
the other edges (which are returned by \procterminscc in such case).
\procterminscc makes use of invariants as well, as they can increase its
precision.
For simplicity we assume that invariants are added
to the input \its 
and that procedure \proccfr
adds such invariants to the refined \its as
well.

The 
effectiveness
of \procterminscc is 
affected
by the kind of ranking functions and invariants used.
Probably the best known kinds of ranking function are \emph{linear ranking functions}~(\lrfs), which have the form
$f(\bar{x}) = a_0+\sum_{i=1}^n a_ix_i$; there are 
efficient algorithms for inferring
them~\cite{PodelskiR04,BagnaraMPZ12}.
However, \lrfs do not suffice for all terminating programs; in such
cases \procterminscc resorts to 
combinations of
 linear
functions, e.g. \emph{lexicographic ranking functions}~(\llrfs)
~\cite{ADFG:2010,Ben-AmramG13jv}.
Apart from performance,  \lrfs have the advantage
that they induce bounds on the length of traces,
and are thus suitable for
complexity analysis.
For invariants, analyzers often use 
abstract domains 
such as
convex polyhedra~\cite{CousotH78}, %
which cannot capture 
more expressive, but expensive disjunctive invariants.
\cfr can
improve termination analysis by 
(1) enabling the inference of 
   \lrfs where without \cfr one would need \llrfs; and
(2) enabling automatic termination proof
   where without \cfr it is not possible.
We suggest
the following schemes for adding \cfr to a  termination analyzer, trading off 
ease of implementation with performance and precision.
\begin{compactitem}
\item (\cfrb): in this scheme \cfr is applied directly to the input
  \its. This is easy to implement but can perform unnecessary refinements that cause overhead in analysis.
\item (\cfrs): in this scheme,
  when \procterminscc fails to prove termination of a \scc, \cfr
  is applied only to the part of the \scc that it could not prove
  terminating.
\item (\cfra): this scheme first collects all edges (from all \sccs)
  on which \procterminscc has failed, and applies \cfr to the input
  \its taking into account only those edges.
\end{compactitem}
For \cfrs and \cfra schemes, \cfr can be applied iteratively
so that refinement is interleaved with termination proofs attempts. In this way, 
each step might introduce further refinements
that could not be done in previous steps. The precision and
performance of these schemes are compared experimentally in
Sect.~\ref{sec:imp}.
Alg.~\ref{alg:termin} shows the pseudocode of a termination
analysis algorithm that uses the above schemes for \cfr. It consists
of two procedures \proctermin and \proctermincfg, and uses
\procterminscc as a black box. It also uses some auxiliary procedures
that we explain below.

\begin{algorithm}[t]
{\small
\DontPrintSemicolon
\setlength{\columnsep}{1cm}
\begin{multicols*}{2}
\Fn{\proctermin{\mbox{\trsys,\cfrb,\cfra,\cfrs}}}{
\lIf{\cfrb}{\trsys = \proccfr{\trsys}}\label{alg:cfrb}  
\emph{F} := \proctermincfg{\trsys}\label{alg:call:1}\;
\While{$\emph{F} \neq \emptyset$ \textbf{and} $\cfra > 0$} {
    $N := \nodes{F}$\label{alg:nodes}\;
    $\trsys := \procremovenrn{\trsys,N}$\label{alg:remnr}\;
    $\trsys := \proccfr{\trsys,N}$\label{alg:cfra}\;
    $\trsys' := \procremovetn{\trsys,N}$\label{alg:remt}\;
    $\cfra := \cfra - 1$\label{alg:redcfra}\;
    $\mathit{F} := \proctermincfg{\ensuremath{\trsys'},\cfrs}$\label{alg:call:2}\;
  }
 \Return{$F==\emptyset$}\label{alg:res:1}
}
\Fn{\proctermincfg{\mbox{\trsys,\cfrs}}}{
$\mathit{F} := \emptyset$\label{alg:init:f};
$\mathit{Q} := [\tuple{\trsys,\cfrs}]$\label{alg:init:p}\;
\While{$\mathit{Q} \neq \emptyset$}{
  $\tuple{\trsys', i} := \mathit{Q.getFirst()}$\label{alg:get}\;
  \ForEach{\emph{\scc} $S$ \emph{of} $\trsys'$} {\label{alg:foreach}
    $F_S := \procterminscc{S}$\label{alg:call:3}\;
      \If {$\mathit{F_S} \neq \emptyset$ and $\mathit{i} > 0$} {
        $\mathit{\trsys''} := \proccfr{\ensuremath{\procbuildits{\ensuremath{F_S}}}}$\label{alg:cfrs}\label{alg:buildits}\;
        $\mathit{Q.add}(\tuple{\trsys'', i - 1})$\label{alg:addtoQ}\;
      }
      \lElse{$F := F \cup \mathit{F_S}$\label{alg:addtoF}}
      }
  }
\Return{$\mathit{F}$}\label{alg:res:2}
}
\end{multicols*}
}
\caption{Pseudocode of Termination Analysis with Control-Flow Refinement}
\label{alg:termin}
\end{algorithm}

\proctermin receives an \its \trsys, a
Boolean \cfrb indicating whether \cfrb should be applied,
and integers \cfra and \cfrs giving the number of times 
that the respective schemes can be applied.
At Line~\ref{alg:cfrb} (L\ref{alg:cfrb} for short) it calls $\proccfr(\trsys)$ if
\cfrb is \textbf{True}, and at
L\ref{alg:call:1} it calls \proctermincfg to analyze the
SCCs of \trsys for the first time. \proctermincfg returns the set $F$
of edges for which it failed to show termination. If
$F= \emptyset$  it halts, otherwise, it enters a loop as
long as  \cfra can be applied ($\cfra>0$) and termination has
not been proven ($F\neq\emptyset$). At L\ref{alg:res:1} it
returns \textbf{True} if $F=\emptyset$, showing that \trsys is
terminating, and \textbf{False} otherwise.

In each iteration of the refine/analyze loop
(L\ref{alg:nodes}-\ref{alg:call:2}) it applies \cfr only to the parts
corresponding to $F$.
At L\ref{alg:nodes} it computes the set of nodes $N$ in
$F$;
at L\ref{alg:remnr} it removes from $\trsys$ nodes that do not reach
nodes in $N$ (since those parts of \trsys have been proven
terminating and they do not affect the \cfr of any node in $N$);
at L\ref{alg:cfra} it applies \cfr considering only loop heads in $N$;
at L\ref{alg:remt} it removes from $\trsys$ nodes not
in $N$ as they have been proven terminating already; at
L\ref{alg:redcfra} it decreases the \cfra scheme counter; and finally
at L\ref{alg:call:2} it calls \proctermincfg again to analyze
$\trsys'$.
Note that the nodes removed at L\ref{alg:remt} are not removed
before at L\ref{alg:remnr} in order to
\begin{inparaenum}[\upshape(a\upshape)]
\item guarantee soundness, as \cfr must consider all possible ways in
  which nodes in $N$ are reached;
\item allow \cfr to benefit from context information; and
\item allow other parts reachable from nodes in $N$ to
  benefit from refinements.
\end{inparaenum}
  
 \proctermincfg analyzes the \sccs of \trsys, and applies \cfr
at the level of \sccs if needed.
At L\ref{alg:init:f} it initializes local variables $F$ and $Q$, where
$F$ is used to accumulate the edges that it fails to prove terminating
(returned at the end at L\ref{alg:res:2}), and $Q$ is a queue of pending (sub)
\itss to be analyzed.
The elements of $Q$ are pairs $\tuple{\trsys',i}$, where $\trsys'$ is
an \its to be analyzed and $i$ is the number of times left to apply
\cfr to its \sccs.
The \emph{while} loop is executed as long as there are pending \itss
in $Q$: at L\ref{alg:get} it takes an \its $\trsys'$ from $Q$, and at
L\ref{alg:foreach}-\ref{alg:addtoF} tries to prove termination of its
\sccs.
This is done by calling \procterminscc on $S$ at
L\ref{alg:call:3}, and if it fails ($F_S\neq\emptyset$) it applies
\cfr at L\ref{alg:buildits}-\ref{alg:addtoQ} to the problematic part
$F_S$ if possible ($i>0$), otherwise it adds $F_S$ to $F$ at
L\ref{alg:addtoF}.
Applying \cfr to $F_S$ is done as follows:
at L\ref{alg:buildits} it builds a new \its from $F_S$, by calling
$\procbuildits{\ensuremath{F_S}}$, and applies \cfr to it.
\procbuildits builds an \its that consists of the nodes and edges of
$F_S$, together with a new entry node that has edges to all nodes of
$F_S$ that are reachable from nodes not in $F_S$. This is because \cfr
must consider all possible ways in which nodes of $F_S$ are reached
(we assume that \procbuildits has global access to the original \its);
and
at L\ref{alg:addtoQ} it adds the refined \its to $Q$ ($\trsys''$ might
have several \sccs now).%

Next we show some examples of how
 Alg.~\ref{alg:termin}, can benefit termination analysis, using
 the different \cfr schemes.
For all examples we give programs and \itss, but we omit
\chcs as they are similar to \itss.
\itss are sometimes simplified (e.g. by joining chains of nodes) for
presentation; however, they are very similar to what we get
in practice -- see Sect.~\ref{sec:imp}.

\cfr can simplify the termination witness from \llrfs to \lrfs, which
is useful if the underlying analyzer does not use \llrfs, and
can help in making cost analysis feasible as well. For example,
consider again method \lst{phases1} of Fig.~\ref{fig:progs}(a) and its
corresponding \its $\trsys$. It is easy to prove that $\trsys$ 
terminates using the \llrf $\tuple{z-y,x}$ for nodes $\node{n_1}$ and
$\node{n_2}$, but it is not possible if we restrict ourselves to
the use of \lrfs. 
Calling $\proctermin(\trsys,\mathit{true},0,0)$ applies \cfr at
L\ref{alg:cfrb} before trying to prove termination, yielding
the \its $\trsys_{pe}$ of Fig.~\ref{fig:progs}(a).
The two loop phases are now explicit: nodes $\node{n_1^2}$ and
$\node{n_2^2}$ correspond to the first phase, and nodes $\node{n_1^1}$
and $\node{n_2^1}$  to the second phase.
Using this refined \its, \proctermincfg finds
\lrfs $z-y$ and $x$ for the corresponding \sccs. 
The next example discusses cases for which \cfr is essential for
proving termination, not only for simplifying the form of the witness.

\begin{example}
\label{ex:termin:search}
Consider method \lst{search} of Fig.~\ref{fig:progs}(b).
It receives an array $q$ representing a circular queue, the size of
the array $n$, the indexes $h$ and $t$ of the head and tail, and a value $v$ to search for.
The loop first (if $h>t$) searches for $v$ in the
interval $[h..n-1]$ and then in $[0..t]$.
This defines two phases where moving from the first to the
second is done %
by setting $h$ to $0$.
It is easy to see that
(a) in the first phase, $n-h$ is non-negative and
decreasing in all iterations except the last, i.e., when setting $h$
to $0$
and
(b) in the second phase, $n-h$ is decreasing
and non-negative.
The last iteration of the first phase makes automatic proofs
subtle, because it breaks the conditions that a \lrf or a
\llrf has to satisfy as function $n-h$ increases when setting $h$ to
$0$.
If we succeed in splitting these phases into separate
loops, such that the last iteration of the first phase is a 
transition that connects them, then proving termination should be
possible with \lrfs only.
Unlike method \lst{phases1} of Fig.~\ref{fig:progs}(a), in
this one the two phases execute the same code, i.e., the
\emph{then} branch at PP\ref{ex:se:inch} (short for program point~\ref{ex:se:inch}).

The \its $\trsys$ of this program is shown in
Fig.~\ref{fig:progs}(b). Node $\node{n_1}$ corresponds to the loop
head, and $\node{n_2}$ to the \emph{if} statement.
Assume that this loop 
is the only loop in a larger program that we cannot prove terminating, so as to take advantage of
applying \cfr at the level of \sccs.
Calling $\proctermin(\trsys,\textit{false},0,1)$ we eventually reach
L\ref{alg:call:3} with \scc $S$ containing nodes \node{n_1} and
\node{n_2}.
$\procterminscc(S)$ fails to prove termination of any edge of $S$ and
returns the set of edges of $S$.
The new \its $\trsys''$ at L\ref{alg:buildits} is like the
\its $\trsys$ of Fig.~\ref{fig:progs}(b), but without the exit node
\node{n_3}.
Applying \cfr at L\ref{alg:cfrs} using properties $\{h=0, h\le t\}$
(or \propscv and \propsc) we obtain $\trsys_{pe}$
 shown in Fig.~\ref{fig:progs}(b), which is added at
L\ref{alg:addtoQ} to $Q$ in order to analyze it again.
Node $\node{n_1}$ of $\trsys$ now has $2$ versions in
$\trsys_{pe}$:
$\node{n_1^2}$ is for the first phase which excludes the last step
that sets $h$ to $0$, which is now handled by the edge
$\edge{\node{n_2^2}}{\varphi_7}{\node{n_1^1}}$; and node
$\node{n_1^1}$ is for the second phase.
When $\trsys_{pe}$ is analyzed, \procterminscc finds \lrfs for both
\sccs of $\trsys_{pe}$.
\ebox
\end{example}

The next example shows (1) how \cfr is useful for inferring precise
invariants, without using disjunctive abstract domains; and (2) 
the importance of the scheme \cfra.

\begin{example}
\label{ex:termin:search_inv}
Let us modify method \lst{search} (and its \its
$\trsys$) to include another  variable $w$, that is initialized
to $t-h+1$ before the loop and is set to $1$ in the \emph{else} branch
at PP\ref{ex:se:else}.
The initial value of $w$ is at least $1$ if $h\le t$, and it is at
most $0$ if $h>t$. However, in the later case the execution eventually
passes through the \emph{else} branch and sets $w$ to $1$. This means
that $w\ge 1$ holds after the loop.
Using the \its $\trsys$ of this program, invariant generators would
fail to infer this information without relying on disjunctive
invariants, e.g. $(h \le t \wedge w\ge 1) \vee (h>t \wedge w\le 0)$
for node $\node{n_1}$. However, when using $\trsys_{pe}$ they succeed
because in
$\trsys_{pe}$ it is explicit that the second phase is reached, in
either way, with $w\ge 1$.
Precise invariants are essential for the precision of termination
analysis. Assume that the loop of method \lst{search} %
is
followed by a second loop ``\lst{while (x>=1)$~$x=x-w;}'', then
termination analysis of this loop would fail without the invariant
$w\ge 1$ since the loop is non-terminating for $w \le 0$.
Note that in order to propagate $w\ge 1$ from the first loop to the
second, we cannot use the scheme \cfrs since it analyzes the \sccs
independently, and thus after applying \cfr to the \scc of the first
loop the new invariants are not propagated to the \scc of the second
loop.
Instead, we can use \cfra by calling
$\proctermin(\trsys,\textit{false},1,0)$.
In this case, the first call to \proctermincfg at L\ref{alg:call:1}
fails to prove any of the two loops terminating, and then \cfr at
L\ref{alg:cfrb} is applied on both loops together and now the
constraint $w\ge 1$ is propagated to the second loop.
One could also use \cfrb, but it might be less efficient if the
program is part of a larger one that does not need \cfr to handle
other \sccs.
\ebox
\end{example}

Next we discuss one of the examples that no tool could handle in the
last termination competition~\cite{termcomp19}, except \rankfinder when using \cfr.

\begin{example}
\label{ex:termin:randomwalk}
Consider method \lst{randomwalk} of Fig.~\ref{fig:progs}(c).  It
simulates a random walk where $w$ is repeatedly increased or decreased at PP\ref{ex:rw:mdw} 
 depending on the random choice
for $c$ at PP\ref{ex:rw:chc}.
The code at PP\ref{ex:rw:if1}-\ref{ex:rw:if4} ensures
termination. %
Assuming that $n \ge 1$,  the loop passes through the following phases:
(a)  $z \ge 1$, so $z$ is decremented at PP\ref{ex:rw:if1}
until it reaches $0$;
(b) since $z=0$ and $i=1$ now, it either executes
PP\ref{ex:rw:if4} and exits the loop, or executes PP\ref{ex:rw:if3}
which decrements $i$ to $0$ and in the next iteration executes
PP\ref{ex:rw:if1} and sets $i$ to $2$ and $z$ back to $n$;
(c) since $z\ge 1$ now, it is decremented at PP\ref{ex:rw:if1} until it
reaches $0$; 
(d) since $z=0$ and $i=2$ now, it either executes PP\ref{ex:rw:if3}
twice, which means that $c=0$ and thus at PP\ref{ex:rw:mdw} $w$ is
decremented twice to $0$ and exits the loop, or it executes
PP\ref{ex:rw:if4} at least once and exits the loop.
The case of $n\le 0$ passes in steps (b) and (d) only.
Applying \cfr to \trsys of Fig.~\ref{fig:progs}(c), e.g. using properties \propscv and \propsdh,
we obtain the \its $\trsys_{pe}$ sketched in Fig.~\ref{fig:progs}(c)
(each box is an \scc with
several nodes and a single cycle that decrements $z$).
Calling \proctermin on $\trsys$ with any of the schemes
refines the graph to something like $\trsys_{pe}$, and then proves
 termination with \lrf $z$ for both \sccs.
\ebox
\end{example}

\paragraph{Cost analysis.} In the rest of this section we discuss the use of \cfr 
for cost analysis, namely the inference of 
upper-bound functions
on the length of traces (see
Sect.~\ref{sec:prelim}).
There are several cost analysis tools for (variants of) \itss, and they are based on similar techniques to those used in
termination analysis.  In particular, they use ranking functions to
infer \emph{visit-bounds}, which are upper-bounds on the number of
visits to edges in the \its.

\koat~\cite{BrockschmidtE0F16} is a tool that works directly on \itss
as defined in Sect.~\ref{sec:prelim} and uses
quasi-\lrfs~\cite{ADFG:2010} to infer visit-bounds. 
\cofloco~\cite{Montoya17} is a tool that works on a form of \itss that
is called cost relations~(\crss), which are similar to recurrence
relations. It mainly uses \lrfs to infer visit-bounds, and it applies
\cfr (directly to \crss) to increase precision. The \cfr techniques of
\cofloco are very similar to those of~\citeN{GulwaniJK09}.
\pubs~\cite{AlbertAGP11} is the first tool to infer upper-bounds for
\crss.
It
uses \lrfs to infer visit-bounds and it does not apply any kind of
\cfr.
In terms of precision, \cofloco can handle anything that \pubs can,
and it is not directly comparable to \koat as there are cases where
one succeed and the other fail to infer upper-bounds.

In the context of cost analysis, \cfr can improve the precision of
inferring visit-bounds in a way that is similar to simplifying the
witness of a termination proof. The next example shows this for all
examples that we have discusses so far. For \cfr in this context
we simply apply procedure \pe of Sect.~\ref{sec:pe} to the input \its.

\begin{example}
\label{ex:cost:overview}
For Fig~\ref{fig:progs}(a), \koat infers a linear
upper-bound without \cfr thanks to the use of quasi-\lrfs,
which allow the inference of $x$ and $z-y$ as
visit-bounds for $\edge{\node{n_2}}{\varphi_4}{\node{n_1}}$ and
$\edge{\node{n_2}}{\varphi_3}{\node{n_1}}$, respectively;
\cofloco infers a linear upper-bound since it applies \cfr that splits
the loop into phases;
\pubs fails to infer any upper-bound, however, it infers a linear
upper-bound when applied to the \its after \cfr.
For Ex.~\ref{ex:termin:search}, \koat and \pubs fail to infer any
upper-bound, but they infer a linear upper-bound after applying \cfr; 
\cofloco infers a linear upper-bound since it
applies its own \cfr that splits the loop into two phases.
For Ex.~\ref{ex:termin:search_inv}, all tools fail to infer an
upper-bound, and all infer a linear upper-bound
after applying \cfr (note that \cofloco's own \cfr
 is insufficient here).
For Ex.~\ref{ex:termin:randomwalk}, all tools fail to infer any
upper-bound, and they infer a linear upper-bound for the refined \its.
\ebox
\end{example}

Next we discuss another direction where ranking functions
are used as properties for \cfr.
Sometimes we can prove termination of an \its but
cannot infer an upper-bound on its cost. One reason is that
 in termination analysis we can use ranking functions that do
not necessarily imply an upper-bound on the length of traces,
e.g. \llrfs.
However, for some kind of \llrfs, even if they do not imply an
upper-bound, they provide useful information about 
loop control flow.
Our goal is to use these ranking functions to help \cfr to 
make this control flow explicit, which helps cost
analysis to infer upper-bounds.

\begin{example}
\label{ex:cost:multiphase}
Consider method \lst{phases2} of Fig~\ref{fig:progs}(d) and its
 \its $\trsys$.
Termination analysis succeeds in proving termination, assigning node
$\node{n_1}$ the \llrf $\tuple{z,y,x}$. This \llrf is in a sub-class called \emph{multi-phase} linear ranking
functions (\mlrfs)~\cite{Ben-AmramG17}; it has the following properties (considering 
consecutive visits to $\node{n_1}$): $z$ decreases,
and when it becomes negative $y$ starts to decrease, and when $y$
becomes negative $x$ starts to decrease.
This means that on reaching $\node{n_1}$ one of the following
must hold: $z \ge 0$, $z<0\wedge y\ge0$, $z<0\wedge y<0\wedge x \ge 0$
or $z<0\wedge y<0\wedge x < 0$.
The \its $\trsys'$ of Fig~\ref{fig:progs}(d) is a 
modification of $\trsys$ making this information explicit:
we added a new node $\node{n_{1a}}$, changed all incoming edges of
$\node{n_1}$ to go to $\node{n_{1a}}$, and added $4$ edges from
$\node{n_{1a}}$ to $\node{n_1}$, each annotated with one
of the above constraints.
Applying \cfr to $\trsys'$ results in the \its $\trsys'_{pe}$ in which
the phases are explicit (nodes $\node{n_{1a}^1}$,
$\node{n_{1a}^2}$, and $\node{n_{1a}^3}$). Cost analysis tools can
infer a linear upper-bound for $\trsys'_{pe}$.
Note that the addition of node $\node{n_{1a}}$ is essential, applying
\cfr directly to $\trsys$ would not do any refinement.
\end{example}

The above example can be generalized as follows. If a loop-head node
$\node{n_i}$ is assigned a \mlrf $\tuple{f_1,\ldots,f_k}$ during
termination analysis: we add a new node $\node{n_{ia}}$ and change all
incoming edges of $\node{n_i}$ to $\node{n_{ia}}$;  and add $k+1$
edges from $\node{n_{ia}}$ to $\node{n_i}$ where the $i$th edge has
the constraints
$f_1(\bar{x}) <0 \wedge \cdots \wedge f_{i-1}(\bar{x}) < 0 \wedge
f_i(\bar{x}) \ge 0 \wedge \nc{\bar{x}}$.
Applying \cfr to the new \its takes advantage of the new edges, and
splits the loop into corresponding phases.

\section{Implementation and Experimental Evaluation}
\label{sec:imp}

In this section we discuss implementation and experimental evaluation
of the ideas discussed in this article. More details are
available in \ref{app:sec:exp} and at the companion website \url{http://irankfinder.loopkiller.com/papers/extra/iclp19}.

We provide a standalone tool that receives an \its
(in several formats) and generates an \its after applying \cfr using partial evaluation~\cite{Gallagher19}.
It accepts any of the properties discussed in Sect.~\ref{sec:pe} (the user selects the heuristics, and then they are inferred automatically), 
as well as user-supplied properties
and can be applied to a subset of the \its.
In addition, it includes an invariants
generator that can be optionally used. %
Apart from the standalone tool, we modified \rankfinder to
incorporate Alg.~\ref{alg:termin}.

We have evaluated the effect of Alg.~\ref{alg:termin} on \irankfinder
using standard sets of benchmarks~\cite{tpdb,Montoya17}.
We analyzed all benchmarks (except those known to be nonterminating) in
two \emph{settings}, using \lrfs and \llrfs, and using different
schemes and properties.
Out of $556$ (resp. $143$) \itss that it cannot handle using \lrfs
(resp. \llrfs) without \cfr, it can handle $251$ (resp. $48$) with
\cfr.
As expected, \cfr has performance overhead that depends on the scheme
and the kind of properties used.
Experiments show that properties $\propsdh$ and $\propsc$ are
important for precision, and that using them with \cfrs
provides the best performance/precision trade-off.
The time spent on \cfr can be up to $42\%$ of the total time,
depending on the scheme and properties used (as expected, \cfrb takes
more time than \cfra and \cfrs).

We have also analyzed this set of benchmarks using other termination
analyzers of ITSs: VeryMax~\cite{BorrallerasBLOR17} and
T2~\cite{BrockschmidtCIK16}. There are 8 (resp. 14) examples that we
can handle due to the use of \cfr and VeryMax (resp. T2) cannot handle
without \cfr.
For benchmarks where VeryMax (resp. T2) failed, we have applied them
again on the \itss after \cfr (using scheme \cfrb) and it could prove
termination of $6$ (resp. $39$) more benchmarks and nontermination of $80$
(resp. $40$) benchmarks.

For cost analysis, we applied \koat on a set of $200$
benchmarks. With \cfr
we got tighter upper-bounds for $15$ benchmarks.
Using \propsdh and \propsc gave the best precision. 
We did not run \pubs on this set since it is not directly designed for
\itss, and more work is required to transform \itss to \crss without
losing precision (i.e., inferring loop summaries which is usually
done by the frontend that uses \pubs).
We did not run \cofloco on this set since it includes a \cfr
component for \crss. However, as we have seen in the examples of
Sect.~\ref{sec:termin}, our \cfr procedure can improve the precision of
both \pubs and \cofloco.
We used the invariant generator of \irankfinder to prove the
 assertions in $13$ programs from~\citeN{SharmaDDA11}.
Without \cfr, it proved the assertions for $5$ of them, and
with \cfr it did so for all benchmarks. Also here,
 \propsdh and \propsc provided the most precise results.
We have also evaluated our \cfr procedure on all examples of Fig.~1
and Fig.~2 of~\citeN{GulwaniJK09}, and got similar refinements.
The tools of~\citeN{SharmaDDA11} and \citeN{GulwaniJK09} are not
available so we cannot experimentally compare to them.

\section{Conclusions, Related and Future Work}
\label{sec:conc}

In this work we
proposed the use of partial evaluation as a general purpose technique
to achieve \cfr.
Our \cfr procedure is developed for \itss, and used a partial
evaluation technique for Horn clause programs based on
specializing programs wrt. a set of supplied
properties. %
The right choice of properties is a key factor for achieving the
desired \cfr, and we suggested several heuristics for
inferring properties automatically.
We provide an implementation that can be used as a preprocessing step
for any static analysis tool that uses \itss, and %
integrated it in \irankfinder in a way that
allows application of \cfr only if needed. %
Experimental evaluation 
demonstrated that our \cfr procedure enables termination proofs and cost inference of many \itss that we could not handle before.

Closely related work has been discussed in Sect.~\ref{sec:intro}. The
use of partial evaluation for \cfr is not directly comparable to these
works; however, experiments show that we achieve similar results. In
particular in the examples of Sect.~\ref{sec:termin} we discussed
some cases that~\citeN{Flores-MontoyaH14} cannot handle, and in
Sect.~\ref{sec:imp} we have seen that we achieve similar refinements
for the examples of~\citeN{GulwaniJK09}. In addition, we could handle
all examples of~\citeN{SharmaDDA11}. 
We conjecture that the partial evaluation algorithm achieves the same refinement as the technique of~\citeN{SharmaDDA11}, provided that the splitter predicates (or their negations) are among the properties provided to the algorithm.  Sharma et al. generate candidate splitter predicates from the conditions occurring in
loops, using the weakest precondition operator to project them onto the loop head, which corresponds closely to our property generation $\propsdh$.

There is a close relationship between partial evaluation of logic programs
(sometimes called partial deduction) and abstract interpretation of logic
program with respect to a goal (top-down abstract interpretation); both
can be expressed in a single generic framework 
parameterised by an abstract domain and an unfolding strategy
\cite{Puebla-Hermenegildo-Gallagher:PEPM99,Leuschel04,PueblaAH06}.
The combination of the two techniques can be mutually beneficial, as shown in
\cite{PueblaAH06}. 
In top-down abstract interpretation, the aim is to derive call- and answer-patterns, which are
described by abstract substitutions over program variables, whereas in partial evaluation, the
aim is to unfold parts of the computation and derive a program specialised for the given goal.
The versions in a polyvariant partial evaluation correspond to multiple call-patterns in
top-down abstract interpretation. Viewing \cfr as an instance of a generic framework, 
the choice of abstract domain and unfolding strategy are crucial, and therefore
this paper focusses on those aspects. There are many ways to achieve polyvariance in partial
evaluation, and obtaining a good set of versions leading to useful refinements requires 
careful choice of the abstract domain,
which in our case is the power set of the given set of
properties.   The control of unfolding is also critical, as unfolding choice predicates leads to
a trade-off between
specialisation and blow-up in the size of the specialised program.

Logic program specialisation has been previously applied as a component in program verification
tools, with a goal similar to the one in this paper.  Namely, the specialisation
of a program with respect to a goal (corresponding to a property to be proved) can
enable the derivation of more precise invariants, contributing to a proof of the property
\cite{Leuschel-Massart-LOPSTR99,DBLP:conf/lopstr/AngelisFPP12,FioravantiPPS12,DBLP:journals/tplp/KafleGGS18}.
Polyvariant specialisation is often crucial, allowing (in effect) the inference of disjunctive
invariants.  Again, for \cfr the choices for abstraction and unfolding strategy are important, to achieve
the right balance between precision and the size of the specialised programs.
In \cite{DBLP:conf/lopstr/AngelisFPP12}, an operation called constrained generalisation is used; this
identifies constraints on a predicate that determine control flow. A generalisation operator on constraints is designed to preserve the control flow. 
This has a relation to 
property-based abstraction but we find it more natural and controllable to let the properties determine the
control flow rather than the other way round.
However, further evaluation of different abstractions is needed.
We performed some experiments on examples from this paper using a general-purpose logic program specialiser (ECCE \cite{LeuschelEVCF06}) but the abstraction used in that tool did not result in any useful \cfr.

For future work, we plan to explore further the automatic inference of
properties, in particular the use of ranking functions as discussed in
Sect.~\ref{sec:termin}. Moreover, we want to explore the use of \cfr
for other static analysis tools, where the data is not numerical.
It is also worth investigating whether the insights from \cfr could be used to improve general-purpose
specialisation tools.
 
\bibliographystyle{acmtrans}

\newpage
\appendix

\section{More details on experiments}
\label{app:sec:exp}

The purpose of this appendix is to provide the reader with extra
information on the experimental evaluation -- all is available at
\url{http://irankfinder.loopkiller.com/papers/extra/iclp19} as well.

For termination analysis, we have experimentally evaluated the effect
of Alg.~\ref{alg:termin} on \irankfinder using $3$ sets of standard
benchmarks:
\begin{itemize}
\item (\bsetA) is a set of $416$ \itss coming from Java programs
  (directory \texttt{From\_Aprove\_2014} of~\citeN{tpdb});
\item (\bsetB) is a set of $806$ \itss coming not necessarily from
  structured programs (directory \texttt{From\_T2} off~\citeN{tpdb});
  and
\item (\bsetC) is a set of $200$ \itss coming mainly from Java and C
  programs used by~\citeN{Montoya17} for experimental evaluation of
  cost analysis.
\end{itemize}
Set \bsetA (resp. \bsetB) includes $159$ (resp. $348$) \its known to
be non-terminating, and thus we exclude them from the experiments.
We first analyzed all benchmarks for termination \emph{without} \cfr
in two \emph{settings}, using \lrfs and using \llrfs, and for those on
which \irankfinder failed, we analyzed them again \emph{with} \cfr
using different schemes and properties.

The results are summarized at Table.~\ref{tab:termin}.
There are $3$ (horizontal) blocks, each corresponding to a set of
benchmarks. On the side of each block, we indicate the total number of
benchmarks on which \irankfinder failed \emph{without} \cfr (excluding
non-terminating ones) and the total time (in minutes) it took to
analyze them, for both \lrfs and \llrfs.
For each set, there are two (vertical) blocks corresponding to the
two settings, using \lrfs and using \llrfs. Columns correspond to
different \cfr schemes, and rows to properties used for \cfr (subsets
of those defined in Sect.~\ref{sec:pe}).
Each entry in the table includes the number of benchmarks that
\irankfinder can now prove terminating and the total time it took to
analyze all benchmarks, using the scheme and properties as indicated
in the corresponding row and column, and the percentage of the total
time spend on \cfr. Each benchmark was given a $5$min time limit.

In terms of precision:
\begin{itemize}
\item For \bsetA: without using \cfr we prove termination of $159$
  (resp. $229$) \itss when using \lrfs (resp. \llrfs), and with \cfr
  we can handle $40$ (resp. $9$) more \itss.
\item For \bsetB: without using \cfr we prove termination of $120$
  (resp. $376$) \itss when using \lrfs (resp. \llrfs), and with \cfr we
  can handle $178$ (resp. $26$) more \itss.
\item For \bsetC: without using \cfr we prove termination of $88$
  (resp. $175$) \itss when using \lrfs (resp. \llrfs), and with \cfr we
  can handle $33$ (resp. $13$) more \itss.
\end{itemize}
In total, we improve for $251$ bechmarks when using \lrfs only, and
$48$ when using \llrfs.
Times in the table should be interpreted carefully, since when
analyzing \emph{without} \cfr and when using \cfrb the analysis stops
as soon as it fails to prove termination of a \scc, while when using
\cfrs and \cfra it keeps trying to prove termination of all \sccs.

For cost analysis, we have applied the tool \koat to estimate the
effect of \cfr on precision. We have analyzed set \bsetC without and
with \cfr with different properties.  \cfr improved (by inferring a
tighter upper-bound) the precision for $15$ benchmarks.

\begin{landscape}

\begin{table}
\caption{Termination benchmarks}
\label{tab:termin}
\begin{minipage}{1.6\textwidth}
\centering

\scalebox{0.95}{%
\begin{tabular}{|c|rrr|rrr|l}

\hline\hline
 & \multicolumn{3}{c|}{LRF} & \multicolumn{3}{c|}{LLRF} & \\
\hline
 Props 
 & \multicolumn{1}{c}{\cfrb} & \multicolumn{1}{c}{\cfrs} & \multicolumn{1}{c|}{\cfra}  & \multicolumn{1}{c}{\cfrb} & \multicolumn{1}{c}{\cfrs} & \multicolumn{1}{c|}{\cfra} &\\
\hline
\hline %
$P_1$
          & 36 (34.36, 22.22\%) %
          & 38 (27.72, 14.51\%) %
          & 39 (32.24, 17.59\%) %
          &  8 (32.15, 7.42\%) %
          &  7 (35.35, 3.72\%) %
          &  9 (32.96, 3.49\%) %
          &\multirow{7}{*}{\rotatebox{-90}{\begin{minipage}{3cm}\scriptsize{\bsetA\\ \lrfs=(106,8.10)\\ \llrfs=(36,14.61)}\end{minipage}}}\\
$P_2$
          & 11 (25.35, 23.41\%) %
          & 15 (23.01, 9.77\%) %
          & 14 (26.75, 13.88\%) %
          &  7 (26.94, 8.81\%) %
          &  4 (32.61, 2.24\%) %
          &  7 (31.78, 2.97\%) %
\\
$P_3$
          & 36 (34.08, 22.27\%) %
          & 38 (27.71, 14.59\%) %
          & 39 (32.21, 17.64\%) %
          &  8 (31.52, 6.67\%) %
          &  7 (35.03, 3.46\%) %
          &  9 (33.64, 3.68\%) %
\\
$P_4$
          & 37 (72.97, 42.16\%) %
          & 39 (44.81, 26.28\%) %
          & 40 (51.64, 30.21\%) %
          &  8 (42.45, 14.72\%) %
          &  7 (38.64, 10.04\%) %
          &  9 (35.99, 7.95\%) %
\\
$P_5$
          & 27 (38.65, 35.86\%) %
          & 28 (28.32, 18.68\%) %
          & 30 (31.95, 21.84\%) %
          &  7 (35.17, 9.20\%) %
          &  4 (37.06, 4.53\%) %
          &  8 (37.84, 4.07\%) %
\\
$P_6$
          & 37 (43.77, 30.39\%) %
          & 39 (33.40, 21.38\%) %
          & 40 (37.95, 23.87\%) %
          &  8 (40.10, 8.89\%) %
          &  7 (40.50, 4.90\%) %
          &  9 (40.75, 4.55\%) %
\\
$P_7$
          & 27 (41.32, 38.83\%) %
          & 28 (28.98, 19.99\%) %
          & 30 (32.94, 23.42\%) %
          &  7 (36.52, 9.60\%) %
          &  4 (37.15, 4.88\%) %
          &  8 (37.86, 4.32\%) %
\\
\hline
\hline %
$P_1$
          & 134 (323.64, 9.15\%) %
          & 153 (313.34, 5.60\%) %
          & 145 (318.82, 5.98\%) %
          &  17 (262.02, 1.05\%) %
          &  14 (271.65, 0.56\%) %
          &  18 (264.59, 0.68\%) %
          &\multirow{7}{*}{\rotatebox{-90}{\begin{minipage}{3cm}\scriptsize{\bsetB\\ \lrfs=(338,159.57)\\ \llrfs=(82,216.96)}\end{minipage}}}\\
$P_2$
          &  28 (267.62, 9.34\%) %
          & 131 (267.23, 3.86\%) %
          &  96 (289.86, 5.21\%) %
          &   3 (255.07, 0.59\%) %
          &   3 (256.33, 0.41\%) %
          &   4 (257.88, 0.36\%) %
\\
$P_3$
          & 134 (322.07, 9.07\%) %
          & 153 (298.56, 5.13\%) %
          & 145 (343.52, 7.00\%) %
          &  17 (262.39, 1.14\%) %
          &  14 (268.12, 0.59\%) %
          &  18 (265.39, 0.64\%) %
\\
$P_4$
          & 139 (397.58, 17.17\%) %
          & 162 (331.17, 12.23\%) %
          & 155 (385.73, 13.94\%) %
          &  22 (252.21, 3.05\%) %
          &  16 (266.69, 1.16\%) %
          &  20 (268.29, 1.63\%) %
\\
$P_5$
          &  63 (333.06,12.54\%) %
          & 151 (306.14, 8.52\%) %
          & 123 (348.52, 8.57\%) %
          &  19 (243.81, 2.55\%) %
          &  13 (254.39, 1.07\%) %
          &  17 (248.11, 1.04\%) %
\\
$P_6$
          & 159 (369.12, 12.08\%) %
          & 178 (334.32, 9.12\%) %
          & 167 (386.57, 9.52\%) %
          &  23 (255.62, 2.42\%) %
          &  19 (261.76, 1.37\%) %
          &  23 (254.75, 1.12\%) %
\\
$P_7$
          &  64 (335.33, 12.72\%) %
          & 150 (307.83, 8.51\%) %
          & 123 (349.56, 9.29\%) %
          &  19 (243.97, 2.65\%) %
          &  13 (255.42, 1.13\%) %
          &  18 (248.11, 1.08\%) %
\\
\hline
\hline %
$P_1$
          & 26 (15.27, 23.67\%) %
          & 21 (14.56, 19.99\%) %
          & 25 (14.35, 20.18\%) %
          &  9 (7.78, 6.25\%) %
          &  7 (9.11, 3.66\%) %
          & 11 (7.55, 6.35\%) %
          &\multirow{7}{*}{\rotatebox{-90}{\begin{minipage}{3cm}\scriptsize{\bsetC\\ \lrfs=(112,4.99)\\ \llrfs=(25,2.30)}\end{minipage}}}\\
$P_2$
          &  4 (11.21, 24.49\%) %
          &  2 (10.60, 16.68\%) %
          &  3 (11.11, 17.53\%) %
          &  6 (4.12, 17.92\%) %
          &  2 (8.67, 2.73\%) %
          &  6 (7.17, 6.18\%) %
\\
$P_3$
          & 25 (15.13, 24.75\%) %
          & 22 (13.18, 20.75\%) %
          & 24 (14.83, 21.19\%) %
          &  9 (7.68, 5.94\%) %
          &  7 (9.12, 3.71\%) %
          & 10 (9.09, 4.39\%) %
\\
$P_4$
          & 30 (25.45, 27.80\%) %
          & 26 (17.33, 29.18\%) %
          & 29 (22.59, 35.19\%) %
          & 12 (5.00, 31.22\%) %
          &  9 (8.69, 5.04\%) %
          & 12 (9.11, 6.15\%) %
\\
$P_5$
          & 20 (14.58, 31.77\%) %
          & 17 (14.45, 29.05\%) %
          & 20 (15.30, 28.09\%) %
          &  9 (3.89, 28.00\%) %
          &  6 (8.07, 12.94\%) %
          & 10 (7.68, 11.45\%) %
\\
$P_6$
          & 33 (21.34, 33.30\%) %
          & 29 (16.65, 29.77\%) %
          & 33 (19.18, 29.75\%) %
          & 12 (4.79, 26.38\%) %
          &  9 (8.63, 14.17\%) %
          & 13 (7.34, 13.01\%) %
\\
$P_7$
          & 21 (15.05, 32.35\%) %
          & 18 (13.27, 28.77\%) %
          & 20 (14.95, 27.28\%) %
          &  9 (3.35, 33.08\%) %
          &  6 (8.78, 5.40\%) %
          & 10 (7.46, 10.07\%) %
\\
\hline\hline
\end{tabular}
}

\vspace{0.2cm}
{\footnotesize$P_1=\{\propsc\}$; $P_2=\{\propsh\}$; $P_3=\{\propsc$,$\propsh\}$;
$P_4=\{\propsc$,$\propscv$,$\propsh$,$\propshv\}$; $P_5=\{\propsdh\}$; $P_6=\{\propsdh$,$\propsc\}$; $P_7=\{\propsdh$,$\propsh\}$}
\end{minipage}

\end{table}
\end{landscape}

\end{document}